\documentclass{aa}  

\usepackage[colorlinks=true]{hyperref}
\hypersetup{linkcolor=blue, citecolor=blue}
\usepackage[draft]{todonotes}
\usepackage{graphicx}
\usepackage{txfonts}
\begin{document} 

\newcommand{\Btot}{$\vec{B_{tot}}$}
\newcommand{\Bord}{$\vec{B_{0}}$}
\newcommand{\deltaB}{$\vec{\delta B}$}
\newcommand{\dangles}{$\delta \theta$}
\newcommand{\CrossTerm}{$\vec{B_{0}} \cdot \vec{\delta B}$}
\newcommand{\MA}{$M_{A}$}
\newcommand{\MS}{$M_{s}$}
\newcommand{\CII}{[C\,{\sc ii}]}
\newcommand{\edcf}{$\rm{\epsilon_{DCF}}$}
\newcommand{\est}{$\rm{\epsilon_{ST}}$}

\newcommand{\kos}[1]{\textcolor{red}{(kostas: #1)}}
\newcommand{\vaso}[1]{\textcolor{violet}{(vaso: #1)}}
\newcommand{\raf}[1]{\textcolor{magenta}{(raf: #1)}}

   \title{Why take the square root? An assessment \\ of interstellar magnetic field strength estimation methods}

   
   
   \author{R. Skalidis  \inst{1, 2}\thanks{\email{rskalidis@physics.uoc.gr}}, J. Sternberg \inst{3}, J. R. Beattie \inst{4}, V. Pavlidou \inst{1, 2} and K. Tassis \inst{1, 2}}

   \institute{
        Department of Physics \& ITCP, University of Crete, GR-70013, Heraklion, Greece
        \and
        Institute of Astrophysics, Foundation for Research and Technology-Hellas, Vasilika Vouton, GR-70013 Heraklion, Greece
        \and 
        Physics Department, Ecole Normale Supérieure, Université  PSL, 24 rue Lhomond, 75005 Paris France
        \and 
        Research School of Astronomy and Astrophysics, Australian National University, Canberra, ACT 2611, Australia}
    
    \authorrunning{Skalidis et al. 2021}
    \titlerunning{Assessment of interstellar B-field strength estimation methods}
    
   \date{}
   

  \abstract
   {The magnetic field strength in interstellar clouds can be estimated indirectly from measurements of dust polarization by assuming that turbulent kinetic energy is comparable to the fluctuating magnetic energy, and using the spread of polarization angles to estimate the latter. The method developed by Davis (1951, Phys. Rev., 81, 890) and by Chandrasekhar and Fermi (1953, ApJ, 118, 1137) (DCF) assumes that incompressible magnetohydrodynamic (MHD) fluctuations induce the observed dispersion of polarization angles, deriving $B\propto 1/\delta \theta$ (or, equivalently, \dangles~$\propto$~\MA, in terms of the Alfv\'{e}nic Mach number). However, observations show that the interstellar medium (ISM) is highly compressible. Recently, Skalidis \& Tassis (2021, A\&A, 647, A186) (ST) relaxed the incompressibility assumption and derived instead $B\propto 1/\sqrt{\delta \theta}$ (equivalently, \dangles~$\propto$~\MA$^2$). 
   }
   %
   {We explored what the correct scaling is in compressible and magnetized turbulence through theoretical arguments, and tested the assumptions and the accuracy of the two methods with numerical simulations.}
   {We used $26$  magnetized, isothermal, ideal-MHD numerical simulations without self-gravity and with different types of forcing. The range of \MA\ and sonic Mach numbers \MS\ explored  are $0.1 \leq M_{A} \leq 2.0$ and $0.5 \leq M_{s} \leq 20$. We created synthetic polarization maps and tested the assumptions and accuracy of the two methods.}
   {The synthetic data have a remarkable consistency with the $\delta \theta \propto M_{A}^{2}$ scaling, which is inferred by ST, while the DCF scaling fails to follow the  data.
   Similarly, the assumption of ST that the turbulent kinetic energy is comparable to the root-mean-square (rms) of the coupling term of the magnetic energy between mean and fluctuating magnetic field is valid within a factor of two for all \MA (with the exception of solenoidally driven simulations at high \MA, where the assumption fails by a factor of 10). In contrast, the assumption of DCF that the turbulent kinetic energy is comparable to the rms of the second-order fluctuating magnetic field term fails by factors of several to hundreds for sub-Alfv\'{e}nic simulations. The ST method shows an accuracy better than $50\%$ over the entire range of \MA\ explored; DCF performs adequately only in the range of \MA\ for which it has been optimized through the use of a ``fudge factor''. For low \MA\ it is inaccurate by factors of tens, since it 
   omits the  magnetic energy coupling term, which is of first order and corresponds to compressible modes. We found no dependence of the accuracy of the two methods on \MS.
   }
   {The assumptions of the ST method reflect better the physical reality in clouds with compressible and magnetized turbulence, and for this reason the method provides a much better estimate of the magnetic field strength over the DCF method. Even in the cases where DCF would outperform ST, the ST method will still provide an adequate estimate of the magnetic field strength, while the reverse is not true. 
   }
   
   
   \maketitle
%
\section{Introduction}

Characterizing the magnetic field properties of the interstellar medium (ISM) is an important task for various fields in astrophysics. Dust polarization is the most widely used magnetic field tracer, since others like the Zeeman effect \citep[e.g][]{heiles_1997, crutcher_2010} or the Goldreich-Kylafis effect \citep{gk_1981, gk_1982} are usually dominated by systematic uncertainties. A major limitation of dust polarization is that it probes directly only the magnetic field morphology and not the strength of the field \citep[e.g.][]{andersson_review}. For this reason, various methods for the indirect estimation of the magnetic field strength from  dust polarization data have been developed. 

\cite{davis_1951} and \cite{chandra_fermi} (DCF) were the first to propose that polarization data can be used to estimate the ISM magnetic field strength. They suggested that magnetic field fluctuations should be imprinted on the polarization map. If there is an ordered magnetic field component which is stronger than the fluctuating one, then the magnetic field lines should be highly ordered, i.e. the dispersion of polarization angles (\dangles) should be low. On the other hand, when fluctuations are stronger or comparable to the mean field component, then magnetic field lines should be highly dispersed and \dangles\ large. They assumed that \dangles\ are due to the propagation of incompressible magnetohydrodynamic (MHD) waves, known as Alfvén waves. The total magnetic energy density of a cloud is,
\begin{equation}
    \label{eq:magnetic_energy}
    \frac{B^{2}}{8\pi} = \frac{1}{8\pi} [B_{0}^{2} + 2 \vec{B_{0}} \cdot \vec{\delta B} + \delta B^{2}],
\end{equation}
where $B_{0}^{2}/(8\pi)$ is the mean field energy density which exists even if there are no fluctuations in the cloud and the other two terms correspond to the fluctuating magnetic field component. Alfvén-wave fluctuations are always perpendicular to \Bord, and thus $\vec{B_{0}} \cdot \vec{\delta B}=0$. DCF further assumed that gas kinetic energy is completely transferred to magnetic energy fluctuations and obtained that,
\begin{equation}
    \label{eq:equipartition_DCF}
    \frac{1}{2} \rho \delta u^2 \approx \frac{\delta B^2}{8\pi}.
\end{equation}
Observationally, \dangles\ is a proxy of $\delta B/B_{0}$ \citep{zweibel_1996}, and hence the above equation can be rearranged in the following form,
\begin{equation}
    \label{eq:dcf_eq}
    B_{0} \approx f \sqrt{4\pi\rho} \frac{\delta u}{\delta \theta},
\end{equation}
where $f$ is a constant factor inserted to account for various biases in the DCF method: only one of three Cartesian velocity components perturbing the field lines \citep{chandra_fermi};  
line-of-sight (LOS) averaging of the polarization signal \citep{zweibel_1990, myers_1991};  averaging within the telescope beam \citep{falceta_2008, houde_2009}, etc. To estimate an appropriate empirical correction for these effects, DCF has been calibrated against numerical magnetohydrodynamic (MHD) simulations \citep{ostriker_2001, heitsch_2001, padoan_2001, Liu_2021} and a factor $f=0.5$ is usually applied to Eq.~(\ref{eq:dcf_eq}).

However, the ISM is highly compressible \citep{heiles_2003} which implies that the DCF approximation of turbulence incompressibility may not be generally applicable. \cite{skalidis_2021} (ST) relaxed this assumption and proposed an alternative method for estimating the magnetic field strength from polarization data in the presence of strongly magnetized and compressible turbulence. They suggested that in compressible but sub/trans-Alfvénic turbulence the coupling term\footnote{We refer to \CrossTerm\ as the coupling term instead of cross-term, since 
\deltaB\ fluctuations are not independent of \Bord.}, \CrossTerm, is not zero and dominates in Eq.~(\ref{eq:magnetic_energy}). As a result, it is the coupling term which should be comparable to the gas kinetic energy rather than $\delta B^{2}/(8\pi)$, and hence,
\begin{equation}
    \label{eq:equipartition_ST}
    \frac{1}{2} \rho \delta u^{2} \approx \frac{\delta B B_{0}}{4\pi}.
\end{equation}
Thus they derived that, 
\begin{equation}
    \label{eq:st_eq}
    B_{0} \approx \sqrt{4\pi\rho} \frac{\delta u}{\sqrt{2 \delta \theta}}.
\end{equation}

Eq.~(\ref{eq:dcf_eq}) and ~(\ref{eq:st_eq}) have a different scaling dependence on \dangles. The strength of \Bord\ in the DCF equation is inversely proportional to  \dangles, while in ST it is inversely proportional to the square root of \dangles.  These scalings can be equivalently expressed as scalings of \dangles\ with the Alfv\'{e}n Mach number $M_A$, which is the ratio of $\delta u$ over the Alfv\'{e}n speed ($V_{A}$).  For DCF,
\begin{equation}
\label{eq:scaling_DCF}
\delta \theta \propto M_{A},
\end{equation}
while for ST,
\begin{equation}
\label{eq:scaling_ST}
    \delta \theta \propto M_{A}^{2}\,.
\end{equation}
This difference is due to the inclusion (ST) or omission (DCF) of the \CrossTerm\ term. \cite{skalidis_2021} compared the performance of Eqs. (\ref{eq:dcf_eq}) and (\ref{eq:st_eq}) against MHD simulations of with $M_A=0.7$, and they showed that Eq. (\ref{eq:st_eq}) is significantly more accurate than Eq. (\ref{eq:dcf_eq}), even when the calibration factor $f$ is included in the latter. However, they did not test the scaling of \dangles\ with $M_A$, as this would require data from a larger number of simulations with varying values of $M_A$. 
The scope of this work is to test the \dangles\ - \MA\ scaling, and thus the validity of DCF and ST assumptions regarding the contribution of \CrossTerm\ in the cloud energetics.

In addition, we address the following theoretical subtlety. Both DCF and ST are statistical methods and represent averaged quantities. From this point of view, one should compare the average kinetic energy density with the average magnetic pressure,
\begin{equation}
    \label{eq:dcf_energy}
    \frac{1}{2} \rho \langle \delta u^{2} \rangle \approx \frac{\langle B^{2} \rangle}{8\pi}.
\end{equation}
For incompressible turbulence, one can directly derive the classical DCF equation, Eq.~(\ref{eq:dcf_eq}), with the substitutions $\delta u \rightarrow \sqrt{\langle \delta u^2 \rangle}$ and $\delta B \rightarrow \sqrt{\langle \delta B^2 \rangle }$, with $\delta \theta$ probing $\sqrt{\langle \delta B^2 \rangle}/B_0$. For compressible turbulence $\vec{B_{0}} \cdot \vec{\delta B} \neq 0$, since it is connected with the density fluctuations \citep{Bhattacharjee_1988, Bhattacharjee_1998}. However, for a periodic signal the number of rarefactions ($\vec{\delta B} \cdot\vec{B_{0}}<0$) is equal to the number of compressions ($\vec{\delta B} \cdot \vec{B_{0}}>0$), and hence $\langle \vec{\delta B} \cdot \vec{B_{0}} \rangle = 0$. This on a first reading implies that DCF (Eq.~\ref{eq:dcf_energy}) also applies to compressible turbulence. 
And yet, the raw DCF method (without an appropriately calibrated $f$ factor) is found to be highly inaccurate when tested in compressible MHD simulations: the estimated \Bord\ values using Eq.~(\ref{eq:dcf_eq}) and $f=1$ systematically deviate from the actual value by large factors. The method produces reasonable estimates only when $f \leq 1/2$ \citep{ostriker_2001, heitsch_2001, padoan_2001, skalidis_2021, Liu_2021}, which means that DCF estimates without a fine-tuned $f$ are more than $100\%$ larger than the actual values. Even when combined with more sophisticated techniques, like the dispersion function analysis \citep{hildebrand_2009, houde_2009, houde_2013} which takes into account the LOS effects, DCF produces highly biased estimates \citep{skalidis_2021}. On the other hand, \cite{skalidis_2021} found that Eq.~(\ref{eq:st_eq}) is significantly more accurate in simulations with compressible turbulence than DCF, whether the latter includes or not the  correction factor $f$. 
Therefore the argument that  the \CrossTerm\ term cancels out in the averaged energetics of strongly magnetized and compressible turbulence does not hold.

In this work, we investigate where the argument fails, and establish that Eq.~(\ref{eq:st_eq}) of ST indeed provides a much superior estimate of the magnetic field strength whenever compressible turbulence is non-negligible, and a good estimate of the magnetic field strength (better than 50\%) across all cases. In summary, we address the following questions: i) What is the actual scaling between \dangles\ and \MA\ in MHD simulations?  ii) Does the \CrossTerm\ term contribute in the cloud energetics when averaging  over the total volume of a cloud? and iii) How accurately can we probe the contribution of \CrossTerm\ with polarization data? 

To address these questions, we significantly expanded the set of numerical tests of both methods (DCF, ST) to a wide variety of forced ideal-MHD, isothermal simulations without self-gravity, with different values of Alfv\'{e}nic Mach numbers $M_A$ (ranging between $0.1$ and $2.0$) and sonic mach number $M_s$ (ranging between $0.5$ and $20$), and with different driving of the turbulence.  We produced synthetic polarization data from all simulations, and we investigated the scaling of \dangles\ with $M_A$, the energetics of the simulations, and the behavior of averaged quantities. Finally, we applied the DCF and ST methods to these data, and we investigated the accuracy of the methods in estimating the strength of \Bord. 

This article is organized as follows. In Sect.~\ref{sec:numerical_simulations} we present the numerical simulations used in this work. In Sect.~\ref{sec:dcf_st_scalings} we test the DCF and ST scaling relations between \dangles\ and $M_A$ in synthetic data. In Sect.~\ref{sec:theory} we show why the omission of \CrossTerm\ in the energetics of compressible turbulence is physically wrong and verify our arguments with numerical data. We also test how accurately we can trace the \CrossTerm\ fluctuations from polarization data. In Sect.~\ref{sec:methods_application} we apply both methods (DCF and ST) in synthetic data and test their accuracy in simulated clouds covering a wide range of \MA\ and \MS. In Sect.~\ref{sec:conclusions} we present the main conclusions of this work.

\section{Numerical simulations}
\label{sec:numerical_simulations}

We used data from the following simulations in our tests. 

"Cho-ENO" \citep{cho_2003_sim, burkhart_2009, portilio_2018, bialy_2020} simulations from the publicly available CATS database \citep{burkhart_CATS_2020}: These are ideal-MHD, isothermal simulations without self-gravity. Turbulence is driven in velocity Fourier space by injecting solenoidal modes only at scales equal to half the size of the simulated cube. Models are characterized by $M_{A}=0.7$ and $2.0$, while \MS\ ranges between $0.7$ and $7.0$. \cite{skalidis_2021} have tested the two methods in the $M_{A}=0.7$ simulations of this dataset, but we also included them in our results for completeness. Simulation data are dimensionless and scale-free. A dimensionless sound speed, which is defined as $\tilde{c_{s}}=\sqrt{\tilde{P}/\tilde{\rho}}$, regulates the units. We assume that the sound speed is $0.91$ km/s for every model and follow \cite{hill_2008} in order to convert to cgs units. The resolution is $256^{3}$. 

"AREPO" simulations from the CATS database \citep{burkhart_CATS_2020}: The simulations setup is presented in detail in \citep{mocz_2017, burkhart_2019}. These are isothermal, ideal-MHD simulations run with the AREPO code \citep{springel_2010}. Turbulence is driven solenoidally until a quasi-static state was reached with $M_{s}=10$ and then self-gravity is switched on. We use the model with $M_{A}=0.35$ at a time step without self-gravity. The resolution of this model is $256^{3}$.

Simulations from \cite{beattie_2020, beattie_2021}. They solve the ideal-MHD equations without self-gravity and isothermal conditions using a modified version of the FLASH code \citep{fryxell_2000, dubey_2008, federrath_2021}. Turbulence is driven in Fourier space by injecting the same amount of power between compressible and solenoidal modes at large scales. Sound speed is $c_{s}=1$ in every model and gas velocities are expressed in \MS\ units. The Alfvénic and sonic Mach numbers cover a wide range of the parameter space, $M_{A}=0.1 - 2.0$ and $M_{s}=0.5 - 20$. Models with $M_{s}=0.5$ have resolution equal to $576^{3}$, while every other model $512^{3}$. 

Simulations of \cite{kortgen_2020}. These are ideal-MHD, isothermal simulations without self-gravity, which were run with the FLASH code \citep{fryxell_2000}. These simulations are in cgs units with $T=11$ K, i.e. sound speed is equal to $0.2$ km/s, and $n=536$ cm$^{-3}$. We use the model with $M_{A}=0.5$ and $M_{s}=7.5$ driven solenoidally. The resolution is $512^{3}$. 

Boundary conditions are periodic in every simulation presented here. In total we used 26 MHD numerical simulations with properties summarized in Table~\ref{table:sims_results}.

\section{Testing DCF and ST scalings with numerical simulations}
\label{sec:dcf_st_scalings}

\begin{figure}
    \centering
    \includegraphics[width=\hsize]{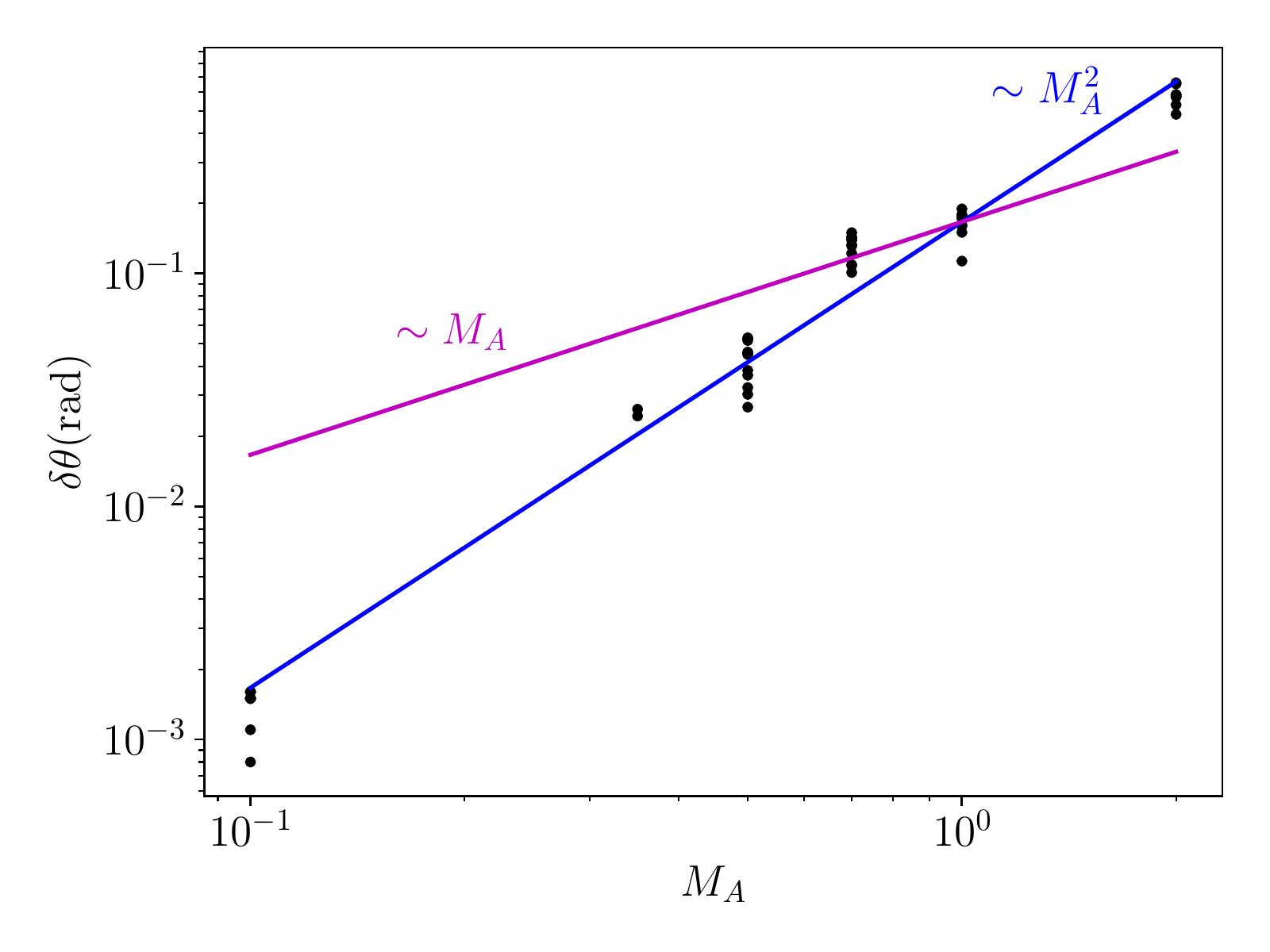}
    \caption{Polarization angle dispersion as a function of the Alfvénic Mach number. Blue line:  ST scaling; magenta line: DCF scaling. The two lines are normalized so that they pass through the data for $M_{A} = 1.0$.}
    \label{fig:dtheta_Ma}
\end{figure}

\cite{skalidis_2021} tested the DCF and ST methods in numerical simulations characterized by a unique \MA\ value (0.7) and five different \MS\ values. However, ISM turbulence spans a wide range of \MA\ and \MS\ values. There is overwhelming observational evidence that ISM turbulence is sub/trans-Alfvénic \citep[e.g.][]{mouschovias_2006, franco_2010, pillai_2015, panopoulou_2016, planck_collaboration_2016, cox_2016, tritsis_2018} and highly compressible (e.g. \citealt{heiles_2003,miville_2007, brunt_2010, burkhart_2015, Orkisz_2017, Nguyen_2019, beattie_2019}). This means that ISM clouds are statistically characterized by $M_{A} \leq 1$ and $M_{s} > 1$. The numerical simulations employed in this work (Sect.~\ref{sec:numerical_simulations}) enable us to test the two methods to a large number of MHD simulations with a wide range of parameters consistent with observations.

The major difference between DCF and ST is the dependence of the magnetic field strength on \dangles. In DCF, $B_{0}$ scales as $\delta \theta^{-1}$ (Eq.~\ref{eq:dcf_eq}) while in ST, $B_{0}$ scales as $\sim \delta \theta^{-1/2}$ (Eq.~\ref{eq:st_eq}). One can divide both equations with $\sqrt{4 \pi \rho}$ and obtain the magnetic field strength in velocity units (the Alfvénic speed, $V_{A}$). The scaling relations of DCF and ST expressed in terms of $V_{A}$ are $V_{A} \sim \delta \theta^{-1}$ and $V_{A} \sim \delta \theta^{-1/2}$, respectively. The Alfvénic speed is,
\begin{equation}
    V_{A} = c_{s}\frac{M_{s}}{M_{A}},
\end{equation}
where $c_{s}$ is the sound speed. Using the above equation with the corresponding \dangles\ scaling dependence of the two methods we obtain,
\begin{equation}
   \label{eq:MA_scaling}
   \delta \theta \propto   
   \begin{cases} 
    M_{A}, & \mbox{\rm{DCF} } \\ 
    M_{A}^{2}, & \mbox{\rm{ST}}
   \end{cases}.
\end{equation}
This is the key difference between the two methods, and it is based on the different scaling relation of \MA\ with the magnetic fluctuations in the incompressible \citep{goldreich_1995} and compressible turbulence \citep{fedderath_2016, beattie_2020}. We tested the two scalings in synthetic polarization data. We computed synthetic $Q$ and $U$ Stokes parameters (Eq.~\ref{eq:Stokes_Q} and ~\ref{eq:Stokes_U} in Sect.~\ref{subsec:Bfield_stokes}) and the polarization angles as $\theta = 0.5\arctan{Q/U}$. Then, we computed the dispersion of the $\theta$ angles, \dangles. All \dangles\ values are shown in Table~\ref{table:sims_results}. 

In Fig.~\ref{fig:dtheta_Ma} we show \dangles\ as a function of \MA\ (we do not show the solenoidally driven models with $M_{A}=2.0$ since their \dangles\ is not representative of the actual fluctuating-to-ordered magnetic field ratio, Sect.~\ref{subsec:app_methods_forcing}). The blue line corresponds to the ST scaling, and the magenta line to the DCF scaling. Both lines are normalized so that they pass through the simulated data for $M_A=1.0$. The ST scaling shows a remarkable consistency with the data. In contrast, the DCF scaling fails to represent the synthetic data. Even if we fine-tune the DCF relation with the use of a factor $f$ \`{a} la \cite{ostriker_2001}, agreement with the data is achieved {\em only around the $M_A$ values used for the tuning}. The scaling slope of DCF is clearly inconsistent with simulation data, regardless of the presence of $f$. 

\section{Theoretical considerations}
\label{sec:theory}

In the previous section we have shown that ST outperforms DCF over a large range of Alfvénic Mach numbers. The simple theoretical arguments to the contrary we discussed in the introduction should, then, be incorrect. In this section we discuss why this is so, from a theoretical perspective. We discuss 

\subsection{Does the $\vec{B_{0}} \cdot \vec{\delta B}$ term contribute to the average energetics of compressible turbulence?}
\label{sec:coupling_energetics}

The coupling term is by definition zero in the incompressible regime and the averaged total energy (kinetic and magnetic) in the perturbations/waves is,
\begin{equation}
    \label{eq:harmonic_energy}
    \langle \delta \epsilon \rangle = \frac{1}{2} \rho_{0} \langle \delta u^{2} \rangle + \frac{\langle \delta B^{2} \rangle}{8\pi}.
\end{equation}
This is similar to the energy equation of a harmonic oscillator, where energy fluctuates between kinetic and potential (here magnetic) forms. The physical analogy in incompressible Alfvénic turbulence works well: the magnetic field oscillates harmonically about \Bord. The lowest value of the potential energy is achieved at the equilibrium state where $\delta B =0$. We can thus consider each fluid element as a harmonic oscillator perturbed around $B_{0}^{2}/(8\pi)$. According to the ergodic theorem, time averaging is equivalent to spatial averaging (equivalently, we can say that within the cloud there exist all possible oscillation phases); Eq.~(\ref{eq:harmonic_energy}) therefore holds - but, in addition, as in the harmonic oscillator, the average values of kinetic and potential energy are equal, $\rho_0\langle \delta u^{2} \rangle/2 = \langle \delta B^{2} \rangle /8\pi$ (Eq.~\ref{eq:equipartition_DCF}), whence DCF Eq.~(\ref{eq:dcf_eq}) is obtained. 

In compressible and strongly magnetized turbulence the fluctuations are also periodic. The $\vec{B_{0}} \cdot \vec{\delta B}$ term, although much higher in absolute value than $\delta B^2$, can be either negative or positive. Therefore  $\langle \vec{B_{0}} \cdot \vec{\delta B} \rangle = 0 $, Eq. (\ref{eq:harmonic_energy}) holds, and it would appear that once again the problem can be reduced to that of a harmonic oscillation. However, in this case the physical analogy is {\em incorrect}. There are two reasons for this. 

First, unlike a harmonic oscillator, the equilibrium state ($\delta B =0$) {\em is not the state of lowest potential energy}; the maximum rarefaction state ($\delta B = -|\delta B|_{max}$) is. The cross term, \CrossTerm, can be either positive or negative and this means that locally it can either add or remove magnetic flux from the fluid elements. When \CrossTerm $>0$, then the mean energy density, $\vec{B_{0}}^{2}/(8\pi)$, locally increases by $2|\vec{B_{0}} \cdot \vec{\delta B}|/(8\pi)$ due to the compression of the magnetic field lines. On the other hand, in regions where  \CrossTerm $<0$ the mean energy density locally decreases by $-2|\vec{B_{0}} \cdot \vec{\delta B}|/(8\pi)$ due to the decompression of the magnetic field lines. In contrast, in a harmonic oscillator (as well as incompressible Alfvénic turbulence), any deviation from the equilibrium position will only increase the potential energy (the magnetic energy, in the case of our fluid elements). 

Second, the dependence of the "potential energy" on the perturbation (here of the magnetic field) is linear: $\delta \epsilon_{\rm p, compressible} \propto \delta B$). In contrast, in a harmonic oscillator the dependence is quadratic: $\delta \epsilon_{\rm p, harmonic} \propto \delta B^2$. Therefore, neither $\delta u$ nor $\delta B$ of a fluid element will behave harmonically with time. 

It is thus clear that the harmonic oscillator is {\em not} the appropriate physical analogue to our problem. Can we substitute it with a more appropriate mechanical analog to guide our intuition?

Indeed we can. Let us consider a shaft with depth $|h|$ below the Earth's surface. We let a basketball fall from height $+h$ above the Earth's surface into the shaft. In classical Newtonian mechanics, the ball will accelerate from a height $+h$ down to the bottom of the hole, $-h$,  where it will bounce upward. In the absence of energy losses, the ball executes the reverse motion as if the clock now runs backwards, and the ball will again reach $+h$, before moving once more downwards towards the bottom of the shaft, continuing these oscillations forever. 
Consider now that we make the choice of taking the zero point of gravitational potential energy to be at the Earth's surface, at its midpoint between its highest value (achieved at $+h$) and lowest value (achieved at $-h$). Now the potential energy $mgz$ is positive above the surface of the Earth ($z>0$), negative below the surface of the earth ($z<0$), and its average value over an entire cycle is $\langle mgz\rangle = 0$. At the same time, the kinetic energy is always non-negative, $\langle mv^2/2 \rangle >0$. It is obvious that in this case, unlike the harmonic oscillator, $\langle mv^2/2 \rangle \neq \langle mgz \rangle$.  Equating them would lead to an absurdity. The absurdity is resolved when we compare the absolute maximum potential energy (i.e., the difference in potential energy between highest and lowest points) with the maximum kinetic energy\footnote{The reader might notice that this is not an exact mechanical equivalent to the case of fluid elements in magnetized compressible turbulence, since in the case of turbulence a second-order term which is always positive is present. To preserve the exact analogy, consider in our mechanical analogue that the ball is attached to a spring which is anchored to the bottom of the shaft, with natural length h, and with spring constant $k$ such that $kh^2/2\ll 2mgh$. The potential energy term is still dominated by $mgz$ and the average kinetic energy will be comparable to $2mgh$; equating it to the average of $kz^2/2$ would still lead to an absurd result, exactly as DCF leads to an incorrect estimate of the ordered magnetic field strength, and an incorrect scaling of \dangles\ with $M_A$ when the \CrossTerm\ term is dominant in absolute magnitude over $\delta B^2$. }: $m v_{\rm max}^2/2 = 2mgh$ yields the correct relation between the maximum positive height $h$, and maximum velocity, $v_{\rm max}$. 

The analogy with \CrossTerm\ oscillations works very well if we replace the basketball with a fluid element and the gravitational potential energy with magnetic energy, where  $\delta B_\parallel$ now plays a role similar to the height of the bouncing ball. The magnetic field of fluid elements oscillates  around \Bord\, just as the ball height fluctuates around $z=0$. For this reason \CrossTerm\ can be positive or negative, i.e. fluctuations in magnetic energy can be positive or negative if its zero point is defined at $\delta B_\parallel=0$ (i.e. $\vec{B}=$\Bord), just as fluctuations in the gravitational potential energy of the ball can be positive or negative if its zero point is defined at $z=0$.  If oscillations are periodic, we obtain $\langle \vec{B_{0}} \cdot \vec{\delta B} \rangle = 0 $. Still, we should not be concluding that the coupling term does not contribute to the average energy budget, any more than we should conclude that the potential energy of the bouncing ball does not contribute to {\it its} average energy budget. 

\subsection{The ST method revisited}

For each fluid element energy oscillates between kinetic and magnetic forms. The total energy density of a fluid element is,
\begin{equation}
    E_{tot} = \frac{1}{2}\rho u^{2} + \frac{B^{2}}{8\pi},
\end{equation}
where $E_{tot} = E_{0} + \delta \epsilon$. In the unperturbed case the total energy of each fluid element is $E_{0} = B_{0}^{2}/(8\pi)$, which is the mean magnetic energy density. When $|\vec{B_{0}}| \gg |\delta B|$, the perturbed energy of each fluid element is,
\begin{equation}
    \label{eq:fluctuating_energy_compr}
    \delta \epsilon \approx \frac{1}{2}\rho \delta u^{2} + 2 \frac{\vec{B_{0}} \cdot \vec{\delta B}}{8\pi},
\end{equation}
in the fluid rest frame where $u_0=0$.
The $\delta B^{2}/8\pi$ term is second order, and hence it was neglected. The $\delta \epsilon$ are energy fluctuations around $E_{0}$ and can be negative when $\delta u \rightarrow 0$ and $\vec{B_{0}} \cdot \vec{\delta B}<0$. If we assume undamped oscillations, then the kinetic energy is completely transferred to magnetic and backwards periodically. It is reasonable to assume that kinetic energy fluctuations are dominated by $\delta u^{2}$ fluctuations when the temperature is constant within a cloud implying $\rho \approx \rho_{0}$. 

To make further progress, we should identify the physically correct way to relate the kinetic energy term with the first-order magnetic energy term.  As in the case of the bouncing ball, the maximum kinetic energy of the fluid element will be comparable to the absolute maximum magnetic energy (i.e., the difference between maximum compression and maximum rarefaction): 
\begin{equation}
\label{eq:STmaxima}
    \rho_0 {\delta u}_{\rm max}^2/2 \sim 2 B_0{\delta B}_{\rm \parallel max} /4\pi. 
\end{equation}
Of course, neither $\delta u_{\rm max}$ nor ${\delta B}_{\rm \parallel max}$ can be probed observationally. The quantities that we do have access to from observations are the spatially-averaged kinetic energy fluctuations, $\langle \delta u ^2 \rangle$ and magnetic field fluctuations  $\sqrt {\langle \delta B^2 \rangle }$ (Sect.~\ref{subsec:Bfield_stokes}), which, by virtue of the ergodic theorem, correspond to the time-averaged fluctuations over an entire period of the evolution. What we need then is a way to relate $\delta u_{\rm max}$ to $\langle \delta u ^2 \rangle$, and $\sqrt {\langle \delta B^2 \rangle }$ to ${\delta B}_{\rm \parallel max}$. 

In the case of the bouncing ball, the time evolution of its velocity $v$ and height $z$ are straightforward to obtain, so we can in fact {\em calculate} these relations between $v_{\rm max}^2$ and $\langle \delta v ^2 \rangle$, and between $z_{\rm max}$ and $\sqrt{\langle z^2\rangle}$. Before we do so, however, and use them in the problem at hand, we should investigate how far we can take the analogy between bouncing ball and fluid element in compressible, strongly magnetized turbulence. Would the time behavior of the fluid element have in fact the same functional form as in the bouncing ball?

The answer is "yes", provided that we can write a formally equivalent Lagrangian for the two systems, and show that the boundary conditions of the problem are similar. If we take the magnetic field perturbation parallel to $\vec{B_0}$, $\delta B_\parallel$, to be a generalized coordinate for the problem, then the generalized velocity would be $\dot{\delta B_\parallel}$, which, by virtue of flux freezing, is proportional to $\delta u_\perp$ (the velocity of the fluid element perpendicular to the magnetic field)\footnote{The result can be obtained, e.g., by differentiation with respect to time of Alfv\'{e}n's theorem.}. In the bouncing ball case, we have a potential energy term that is proportional to $z$, and a  kinetic energy term that is proportional to $\dot{z}^2$. Similarly, for the fluid element responsible for a magnetic field compression or rarefaction, we have a potential energy term that is proportional to $\delta B_{\parallel}$, and a  kinetic energy term that is proportional to $\delta u_\perp^2\propto \dot{\delta B_{\parallel}}^2$.   The Lagrangians of the two problems are thus formally equivalent. The boundary conditions of the problem are also similar: the presence of the bulk medium and its large-scale magnetic field acts as the ``ground'', forcing the fluid element with increasing velocity undergoing a rarefaction to reverse course back towards increasing magnetic field with velocity decreasing in magnitude. As a result, the time evolution profiles of $\delta B_{\parallel}$ and $\dot{\delta B_{\parallel}}$ will be similar to those of $z$ and $v$, respectively, in the bouncing ball problem. 

For the bouncing ball,  over one period of the motion  
$\langle v ^2 \rangle = v_{\rm max}^2/3$ and $\langle z^2 \rangle = 7h^2/15$. Therefore, by eliminating $v_{\rm max}^2$ in favor of $\langle v ^2 \rangle$, and $h$ in favor of $\sqrt{\langle z^2 \rangle}$, the relation between maximum kinetic and potential energies, $mv_{\rm max}^2/2 = 2mgh$, can be rewritten as
\begin{equation}
\label{eq:exactaverage}
\frac{1}{2}m \langle v ^2 \rangle =\frac{2}{3}\sqrt{\frac{15}{7}}mg \sqrt{\langle z^2 \rangle} 
\approx 0.98 mg \sqrt{\langle z^2 \rangle}\,.
\end{equation}
Since the dynamics of the two systems are analogous, we can also express Eq.~(\ref{eq:STmaxima}) for compressible fluctuations as\footnote{We note that the $\langle u_{\perp}^2\rangle$ that we probe observationally refers to the LOS component of the velocity, since velocity dispersions are obtained from the Doppler broadening of emission lines. Since polarization measurements only probe the plane-of-the-sky (POS) component of the magnetic field, the LOS velocity is indeed the perpendicular component, as in the Lagrangian analogy between bouncing ball and fluid element in strongly magnetized compressible turbulence.}, 
\begin{equation}
    \label{eq:energy_equillibrium_st}
\rho_0 \langle  u_{\perp}^2\rangle /2 \sim B_0 \sqrt{\langle \delta B^2_{\parallel}\rangle}/4\pi\,,
\end{equation}
from which we derive,
\begin{equation}
    \label{eq:st_model}
    B_{0} \approx \sqrt{4 \pi \rho_{0}} 
    \sqrt{\langle u_{\perp}^{2} \rangle} \left[ 2\frac{
    \sqrt{\langle \delta B_{\parallel}^{2} \rangle}}{ B_{0}} \right]^{-1/2}.
\end{equation}

\subsection{Comments on the assumptions and approximations of the ST method}
\label{subsec:comments_st_method}

In this section we discuss some assumptions entering the ST method which can limit its accuracy.  
Firstly, we assumed that oscillations are undamped in compressible turbulence, but in reality shocks lead to significant energy loss and non-ideal effects (e.g. ambipolar diffusion) can induce significant loss of magnetic flux from the cloud. Even in ideal-MHD simulations there is significant energy dissipation due to the presence of shocks and numerical diffusion. In these cases, equipartition between the root-mean-square (rms) kinetic and magnetic energy is not guaranteed. In nature, equipartition can hold when there is a constant energy source at large scales injecting energy to the cloud. In numerical simulations this is achieved with the so called forcing, applied in most cases. Secondly, we assumed that magnetic pressure dominates over gas pressure. This is a reasonable approximation for clouds with $M_{A}<1$ and $M_{S}>1$, but when $\vec{B_{0}} \cdot \vec{\delta B}<0$ the magnetic pressure may locally become comparable to the gas pressure (especially in cases when \CrossTerm\ is close to its negative minimum). However, an oscillator spends only a limited amount of time at its negative minimum, which through ergodicity means that gas pressure will be important only for a small volume fraction of the cloud. Thirdly, to arrive to  Eq.~(\ref{eq:energy_equillibrium_st}) we assumed that gas density fluctuations are much smaller than $u^{2}$, and hence that density is approximately constant in the integral, $\rho \approx \rho_{0}$. Thus, Eq.~(\ref{eq:st_model}) omits any contribution from density fluctuations in the kinetic energy, $\langle \delta \rho u^{2} \rangle/2$. 

\begin{figure}
        \centering
        \includegraphics[width=\hsize]{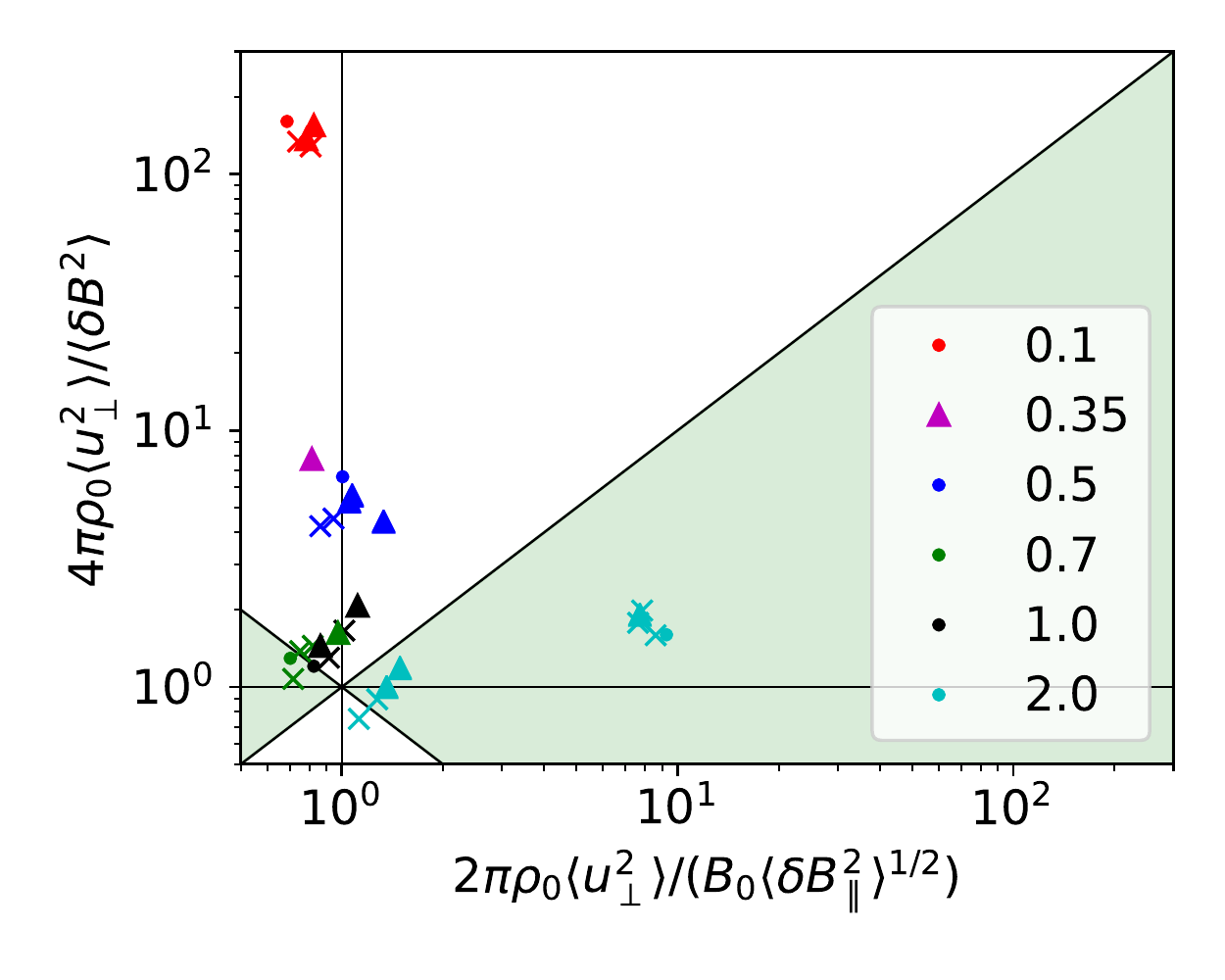}
        \caption{Relative ratio of kinetic over magnetic energy density using Eq.~(\ref{eq:dcf_energy}) (vertical axis) and Eq.~(\ref{eq:energy_equillibrium_st}) (horizontal axis). Colors correspond to different \MA\ as shown in the legend. Dots correspond to simulations with $M_{s}<1$, "x" to $1 < M_{s} \leq 4$ and triangles to $M_{s}>4$.}
     \label{fig:energetics_ratios}
\end{figure}
\begin{figure}
    \centering
    \includegraphics[width=\hsize]{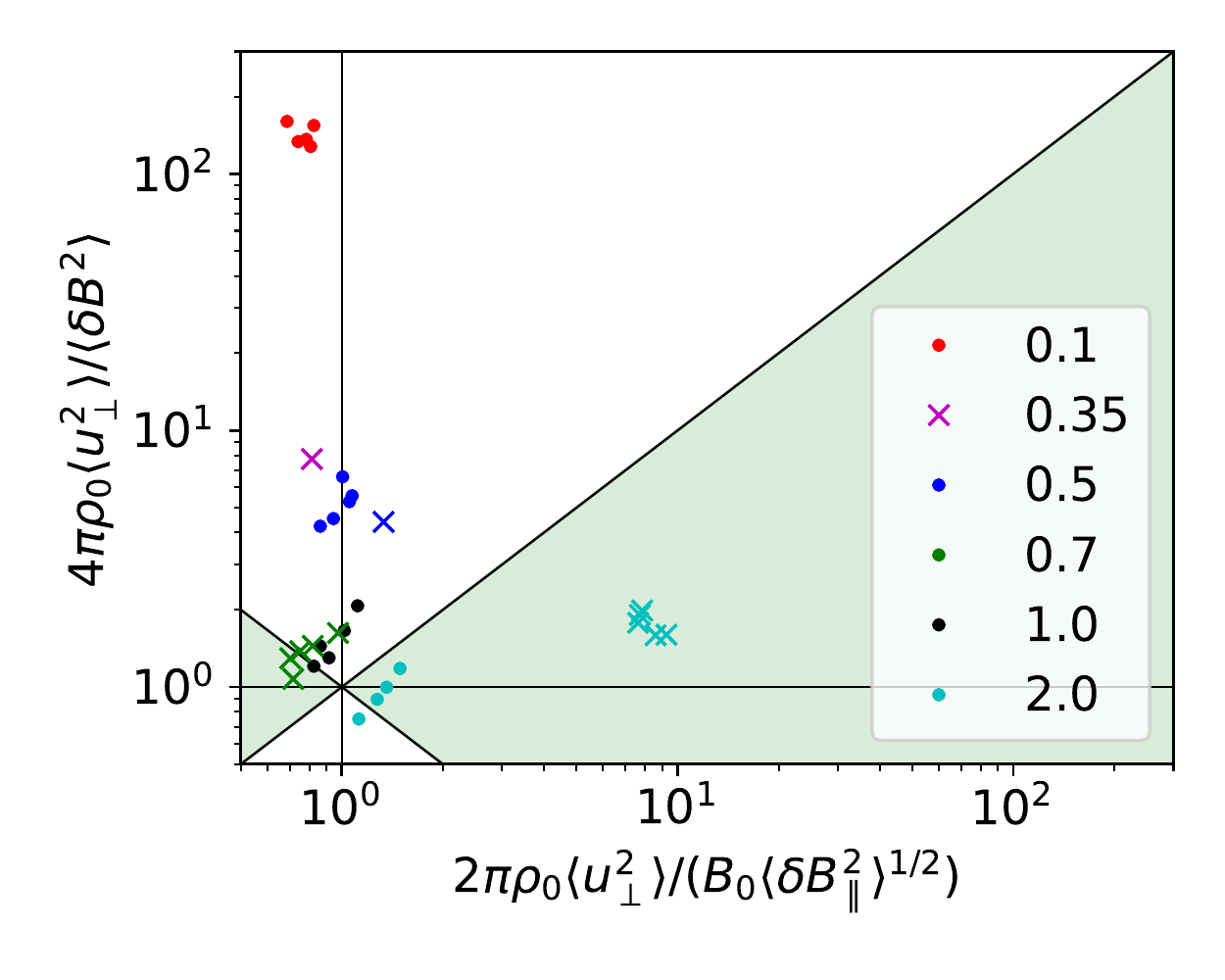}
    \caption{Same as in Fig.\ref{fig:energetics_ratios}. Dots correspond to models driven with a mixture of compressible and solenoidal modes and "x" denotes models driven solenoidally.}
    \label{fig:energetics_ratios_forcing}
\end{figure}

\subsection{How good are the energy equipartition assumptions of DCF and ST?}
\label{sec:testing_theory}

In order to assess the validity of Eqs.~(\ref{eq:dcf_energy}) and ~(\ref{eq:energy_equillibrium_st}), we tested them in numerical simulations. We computed the kinetic energy term ($\rho_0 \langle u_\perp^2 \rangle/2$) and compared it against the two different magnetic energy terms. 

In Fig.~\ref{fig:energetics_ratios} we show in the vertical axis the ratio of kinetic over  magnetic energy, with both terms calculated from Eq.~(\ref{eq:dcf_energy}) (DCF). In the horizontal axis, we show the ratio of kinetic over magnetic energy, with both terms calculated from Eq.~(\ref{eq:energy_equillibrium_st}) (ST). Different color points correspond to simulations with different \MA, while different shapes correspond to simulations with different \MS. Subsonic ($M_{s}<1$) simulation results are shown with dots, supersonic models with $1 < M_{s} \leq 4$ are shown with an "x" and $M_{s}>4$ models are shown with triangles. The vertical line indicates exact equipartition \`{a} la ST and the horizontal line exact equipartition \`{a} la DCF. Diagonal lines separate the regions where each method outperforms the other. Green-shaded regions are closer to the horizontal than the vertical line; there, it is the quadratic term in magnetic energy that dominates and is better comparable to the kinetic energy. White-shaded regions are closer to the vertical line; there, it is the coupling term in the magnetic energy that dominates, and it is that term that is closer to the kinetic energy. 

There is a strong dependence of these results on \MA. The coupling term dominates the energetics of sub-Alfvénic simulations. This has been already shown in numerical simulations of strongly magnetized and compressible turbulence \citep{fedderath_2016, beattie_2020}. There is a slight difference between Eq.~(\ref{eq:energy_equillibrium_st}) and the one from \cite{fedderath_2016} and \cite{beattie_2020}. We include only fluctuations which are parallel to \Bord, while they included the total \deltaB\ rms, and we include only the perpendicular velocity rms in the kinetic energy, while they included the total velocity rms; however, the overall behaviour is the same. 

Our results suggest that Eq.~(\ref{eq:dcf_energy}) (DCF) is highly inaccurate (by factors of several to 100) in sub-Alfv\'{e}nic simulations, performs comparably to Eq.~(\ref{eq:energy_equillibrium_st}) (ST) in trans-Alfv\'{e}nic simulations, and clearly outperforms Eq.~(\ref{eq:energy_equillibrium_st}) (ST) only in trans/super-Alfvénic turbulence simulations driven solenoidally (see Fig.~\ref{fig:energetics_ratios_forcing} and also discussion in Sect.~\ref{subsec:testing_theory_forcing}). This shows the weakness of the incompressible approximation to accurately describe the energetics of sub-Alfvénic turbulence and indicates that the DCF method is precarious to use when the  $M_{A}$ of a cloud is unknown. On the other hand, ST is fairly accurate (better than factor of 2) in the entire sub/trans-Alfvénic regime, even for high \MS\ cases. Overall, the coupling term dominates in the energetics over the $\delta B^{2}$ term when $M_{A}<1$ and cannot be ignored. This is consistent with our discussion in Sect.~\ref{sec:theory} and highlights the importance of this term when estimating the magnetic field strength.

\subsection{How does the forcing affect the energetics?}
\label{subsec:testing_theory_forcing}

Turbulence in MHD simulations is driven in order to achieve the desired \MS. Driving is implemented by injecting compressible or incompressible or a mixture of modes in the cloud through a stochastic process in Fourier space. This process is supposed to mimic the driving mechanisms found in nature. We explored if there is any dependence of the results shown in Fig.~\ref{fig:energetics_ratios} on the forcing mechanism.

In Fig.~\ref{fig:energetics_ratios_forcing} we show the same data as in Fig.~\ref{fig:energetics_ratios}, but here dots correspond to models driven with a mixture of compressible and incompressible modes and "x" represents solenoidally driven simulations. The driving mechanism of each simulation is shown in the second column of Table~\ref{table:sims_results}. The majority of simulations used in the current work were driven with an equal mixture of modes, but there are also a few driven solenoidally. Overall, our results in Fig.~\ref{fig:energetics_ratios_forcing} are weakly affected by the forcing mechanism in the sub-Alfvénic simulations. 

We compare the effect of forcing at models with $M_{A}=0.5$ and $M_{A}=2.0$, since we have an overlap of both solenoidally and mixed driven simulations. Blue points correspond to models with $M_{A}=0.5$. The solenoidally driven model (shown with the blue "x") has the largest kinetic over $B_{0}\langle \delta B_{\parallel}^{2} \rangle^{1/2}$ ratio. The reason is that forcing injects only incompressible modes in the cloud which are only traced by $\delta B^{2}$. On the other hand, when mixed forcing is used the kinetic energy is equally shared among $B_{0}\langle \delta B_{\parallel}^{2} \rangle^{1/2}$ and $\delta B^{2}$. For this reason, the blue "x" point is shifted towards larger and smaller values in the horizontal and vertical axis respectively compared to the blue dots. However, the difference between the forcing mechanisms does not create significant deviations between these models. In sub-Alfvénic turbulence \Bord\ is much stronger than the forced perturbations and determines how energy is transferred among the modes.

The effect of forcing in super-Alfvénic simulations is more prominent. We compare the models with $M_{A}=2.0$ (shown with cyan) and mixed forcing (denoted by dots) with the solenoidally driven models (denoted by $\times$). Models with mixed forcing are clustered in the bottom left corner of Fig.~\ref{fig:energetics_ratios_forcing}, while models with solenoidal forcing in the right bottom corner. The reason is that solenoidal modes are not represented by $\delta B_{x}$, and hence all the injected kinetic energy goes to $\delta B^{2}/(8\pi)$. On the other hand, when mixed forcing is applied the injected energy is shared. In super-Alfvénic turbulence $|\vec{B_{0}}| < |\vec{\delta B}|$, and hence the dynamics of the cloud are determined by \deltaB, instead of \Bord. In these cases the mode of the forced fluctuations determines the cloud dynamics. Thus, there is a large separation between the solenoidal and mixed driven simulations with $M_{A}=2.0$ in Fig.~\ref{fig:energetics_ratios_forcing}. Magnetic field lines are highly curved in the solenoidally driven simulation, and hence $\delta B^{2}/(8\pi) \gg B_{0}\langle \delta B_{\parallel}^{2} \rangle^{1/2}$. On the other hand, in mixed driven simulations $\delta B^{2}/(8\pi) \approx 2 B_{0}\langle \delta B_{\parallel}^{2} \rangle^{1/2}$. This is in striking contrast with the sub-Alfvénic turbulence where fluctuations evolve independently of the forcing mechanism and fluctuations are dictated by \Bord\ and not by \deltaB. 

\subsection{Can we trace \CrossTerm\ with polarization data?}
\label{subsec:Bfield_stokes}

The DCF and ST methods can be used to estimate the strength of \Bord\ using the incompressible and compressible approximation respectively, provided that we can estimate from observations $\rho_0$, $\langle \delta u_\perp^2 \rangle $, and $\langle \delta B_\perp^2\rangle$ (for DCF) or $\langle  \delta B_\parallel^2 \rangle$ (for ST). 

Observationally, the gas volume density is one of the most uncertain parameters inserted in these equations. In the diffuse ISM, $\rho_{0}$ can be estimated using specific gas tracers, like the $158 \mu$m fine structure transition line of \CII\ \citep[e.g.][]{langer_2010, goldsmith_2018}. 
The rms velocity can be estimated using the turbulent broadening of the dominant gas tracer at each cloud and the fluctuating-to-ordered magnetic field ratio from the dispersion of the polarization angle data, \dangles.

One concern raised about the ST method is that \dangles\ is considered to be tracing only  fluctuations perpendicular to the mean magnetic field \citep[e.g.][]{zweibel_1996} and, as such, is dominated by the Alfvénic modes. However, in the ST method (Eq.~\ref{eq:st_model}) it is the $\langle \delta B_{\parallel}^{2} \rangle^{1/2}/B_{0}$ term that is inserted in the energetics, rather than  $\langle \delta B_{\perp}^{2} \rangle^{1/2}/B_{0}$. In this section we explore if this indeed poses a problem for the ST method, or whether we can indeed probe parallel magnetic fluctuations from \dangles. 

Consider an ISM cloud permeated initially by an undisturbed and homogeneous magnetic field $\vec{B_{0}} = (B_{0}, 0, 0)$. We perturb \Bord\ with  $\vec{\delta B} = (\delta B_{x}, \delta B_{y}, \delta B_{z})$ and we assume that \deltaB\ is random, and hence $\langle \vec{\delta B} \rangle = 0$. The total magnetic field is $\vec{B_{tot}} = \vec{B_{0}} + \vec{\delta B}$, 
\begin{equation}
    \label{eq:Btot_components}
    \vec{B_{tot}} = (B_{x}, B_{y}, B_{z}) = (B_{0} \pm \delta B_{x}, \pm \delta B_{y}, \pm \delta B_{z}).
\end{equation}
If the LOS is parallel to the z-axis, then the projected magnetic field morphology of \Btot\ as traced by dust polarization will be given by the Stokes parameters \citep{lee_draine_1985},
\begin{equation}
    \label{eq:Stokes_I}
    I(x, y) = \int \rho dz,  
\end{equation}
\begin{equation}
    \label{eq:Stokes_Q}
    Q(x, y) = \int \rho \frac{B_{x}^{2}-B_{y}^{2}}{|B|^{2}} d z,
\end{equation}
\begin{equation}
    \label{eq:Stokes_U}
    U(x, y) = 2 \int  \rho \frac{B_{x} B_{y}}{|B|^{2}} d z,
\end{equation}
where $|B|$ denotes the total strength of the field. The above equations hold when dust grain properties are uniform and temperature is constant throughout the cloud. For simplicity we assume that density is uniform along each LOS and that $|B|^{2}$ variations along the LOS are negligible (i.e. that the $\rho_0$ and $B_0^2$ terms strongly dominate $\rho$ and $|B|^2$, respectively, and thus the latter can be pulled out of the integrals). From Eq.~(\ref{eq:Btot_components}) and ~(\ref{eq:Stokes_Q}) we derive for the $Q$ Stokes parameter,
\begin{equation}
    \label{eq:Stokes_Q_expanded}
    Q(x, y) \approx \rho_0 B_0^{-2}\int \left ( B_{0}^{2} \pm 2 B_{0} \delta B_{x}+\tilde{\delta B}^{2} \right ) dz,
\end{equation}
where $\tilde{\delta B}^{2} = \delta B_{x}^{2} - \delta B_{y}^{2}$. Using Eq.~(\ref{eq:Btot_components}) and~(\ref{eq:Stokes_U}) the $U$ Stokes parameter can be expressed as,
\begin{equation}
    \label{eq:Stokes_U_expanded}
    U(x, y) \approx 2\rho_0 B_0^{-2}\int \left ( \pm B_{0} \delta B_{y} \pm \delta B_{x} \delta B_{y} \right ) dz \,.   
\end{equation}

\begin{figure}
        \centering
        \includegraphics[width=\hsize]{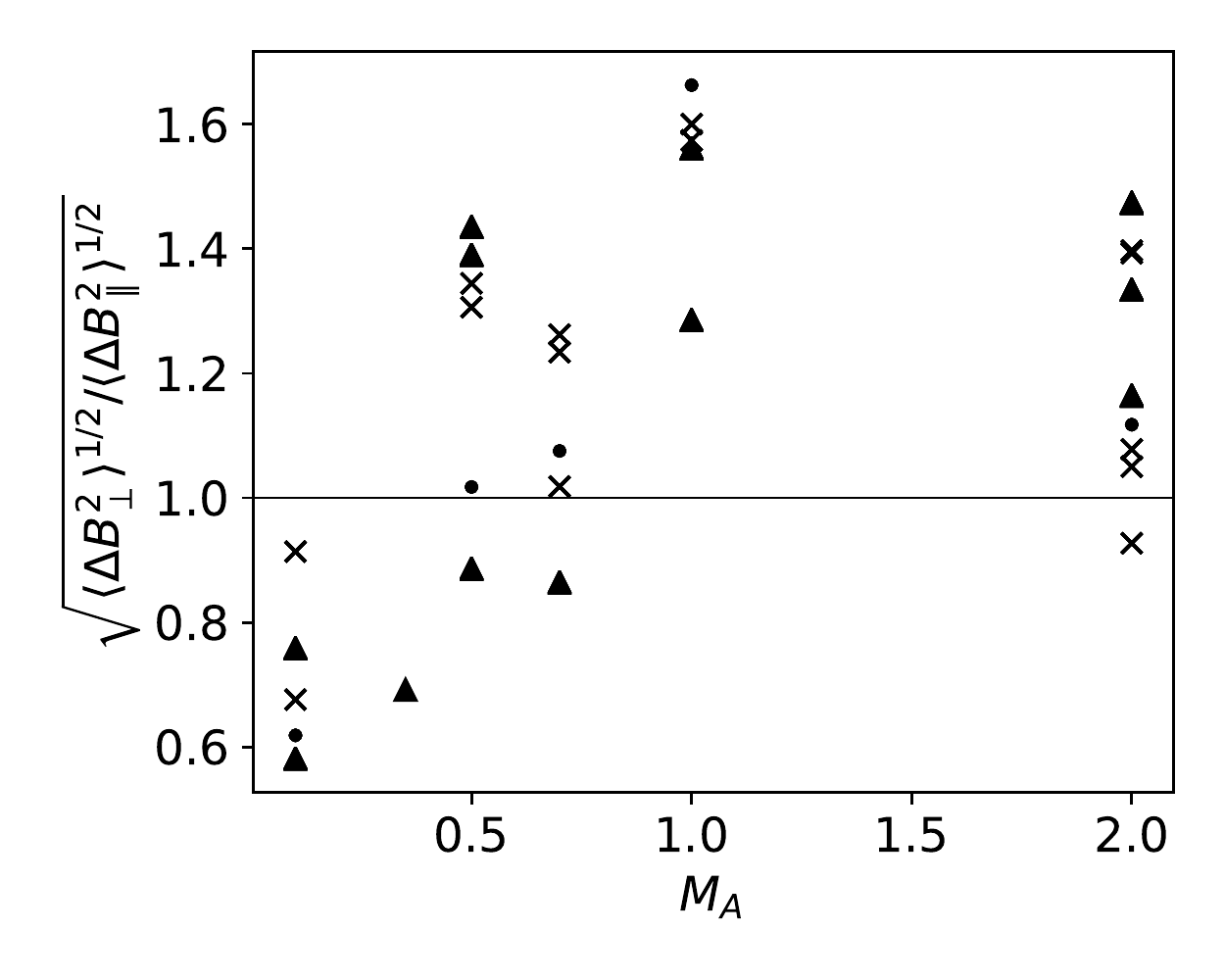}
        \caption{Ratio of perpendicular over parallel LOS averaged magnetic field fluctuations.} 
     \label{fig:dtheta_assumptions}
\end{figure}

If  $|\vec{B_{0}}| \gg |\delta B|$, then we can drop the second-order terms, obtaining, 
\begin{equation}
    \label{eq:Q_stokes_approx}
    Q(x, y) \approx \rho_0 B_0^{-2} \int \left ( B_{0}^{2} \pm 2 B_{0} \delta B_{x} \right ) dz , 
\end{equation}
\begin{equation}
    \label{eq:U_stokes_approx}
    U(x, y) \approx \rho_0 B_0^{-2} \int 2\left ( \pm B_{0} \delta B_{y} \right ) dz .    
\end{equation}
The polarization angle is $\tan{2\theta} = U/Q$. In the $|\vec{B_{0}}| \gg |\delta B|$ regime $\theta$ is small and $\tan{2\theta} \approx 2\theta$. From Eq.~(\ref{eq:Q_stokes_approx}) and ~(\ref{eq:U_stokes_approx}) we obtain,
\begin{equation}
    \label{eq:theta}
    \theta \approx \frac{U}{2Q} \approx \frac{ \int \left ( \pm \delta B_{y} \right ) dz}{\int \left ( B_{0} \pm 2 \delta B_{x} \right ) dz }.   
\end{equation}
The $\delta B_{x}$ term in the denominator of Eq.~(\ref{eq:theta}) is due to turbulence compressibility. In the limit where perturbations are limited only to Alfv\'{e}n waves, this term is by definition zero and $\theta$ traces the perpendicular fluctuations of the magnetic field fluctuations \citep{zweibel_1996}. 

The dispersion of polarization angles is $\delta \theta^{2} = \langle \theta^{2} \rangle_{2D} - \langle \theta \rangle_{2D}^{2} =  \langle \theta^{2} \rangle_{2D}$, since $\langle \theta \rangle_{2D} = 0$, where brackets here denote averaging in the x-y plane. For convenience we adopt the following notation: 
\begin{align}
    & \Delta B_{y} = \int \left ( \pm \delta B_{y} \right ) dz, \\ 
    & \Delta B_{x} = \int \left ( \pm \delta B_{x} \right ) dz, \\ 
    & \tilde{B_{0}} = \int  B_{0}  dz.
\end{align}
Thus, we obtain,
\begin{equation}
\label{eq:firstdispersion}
    \delta \theta = \left \langle \frac{\Delta B_{y}^{2}}{\tilde{B_{0}}^{2} + 4 \tilde{B_{0}} \Delta B_{x} + 4 \Delta B_{x}^{2}} \right\rangle_{2D}^{1/2}.
\end{equation}
Since $|\vec{B_{0}}| \gg |\delta B|$, $4 \Delta B_{x}^{2}$ is of second order, and hence \dangles\ is simplified to, 
\begin{equation}
    \label{eq:dtheta_general}
    \delta \theta \approx \left \langle \frac{\Delta B_{y}^{2}}{\tilde{B_{0}}^{2} + 4 \tilde{B_{0}} \Delta B_{x}} \right\rangle_{2D}^{1 / 2}.
\end{equation}
The $\tilde{B_{0}} \Delta B_{x}$ term is due to $\delta B_{x}$ and represents the average coupling of \Bord\ with \deltaB. 

It is therefore clear that although $\Delta B_{x}$ does contribute to the dispersion of $\delta \theta$, this contribution comes from a first-order term in the denominator of the average in the RHS of Eq.~(\ref{eq:dtheta_general}). When $|\vec{B_{0}}| \gg |\vec{\delta B}|$, parallel fluctuations have but a limited contribution in \dangles\ and can be neglected. Thus, to first order, \dangles\ represents perpendicular fluctuations: $\delta \theta \sim \sqrt{\langle \Delta B_y^2 \rangle}/\tilde{B_{0}}$. 

However, in the ST method, it is parallel fluctuations that need to be estimated and used in  (Eq.~\ref{eq:st_model}). Only when $\langle \Delta B_{y}^{2}\rangle_{2D} / \tilde{B_{0}} \sim \langle \Delta B_{x}^{2} \rangle _{2D} / \tilde{B_{0}}$ would \dangles\ be an adequate metric of parallel fluctuations. 
The question then becomes: do parallel and perpendicular fluctuations have similar dispersions in compressible turbulence?  

The answer is ``yes'', as we show in Fig.~\ref{fig:dtheta_assumptions}, where we have plotted the $\sqrt{\langle \Delta B_{\perp}^{2}\rangle_{2D} / \langle \Delta B_{\parallel}^{2} \rangle _{2D}}$ ratio as a function of \MA. We find that in all cases the dispersion of perpendicular fluctuations is comparable to that of parallel fluctuations, with deviations always smaller than a factor of 2. These results are consistent with the findings of \citet{beattie_2020}. Note that in Fig.~\ref{fig:dtheta_assumptions} we are displaying the square root of the dispersions of the LOS averages of perpendicular and parallel fluctuations, respectively, since these are the relevant quantities in relating our observable ($\delta \theta$) with the quantity of interest for the ST method ($\delta B_\parallel$).
\begin{figure*}
        \centering
        \includegraphics[width=\hsize]{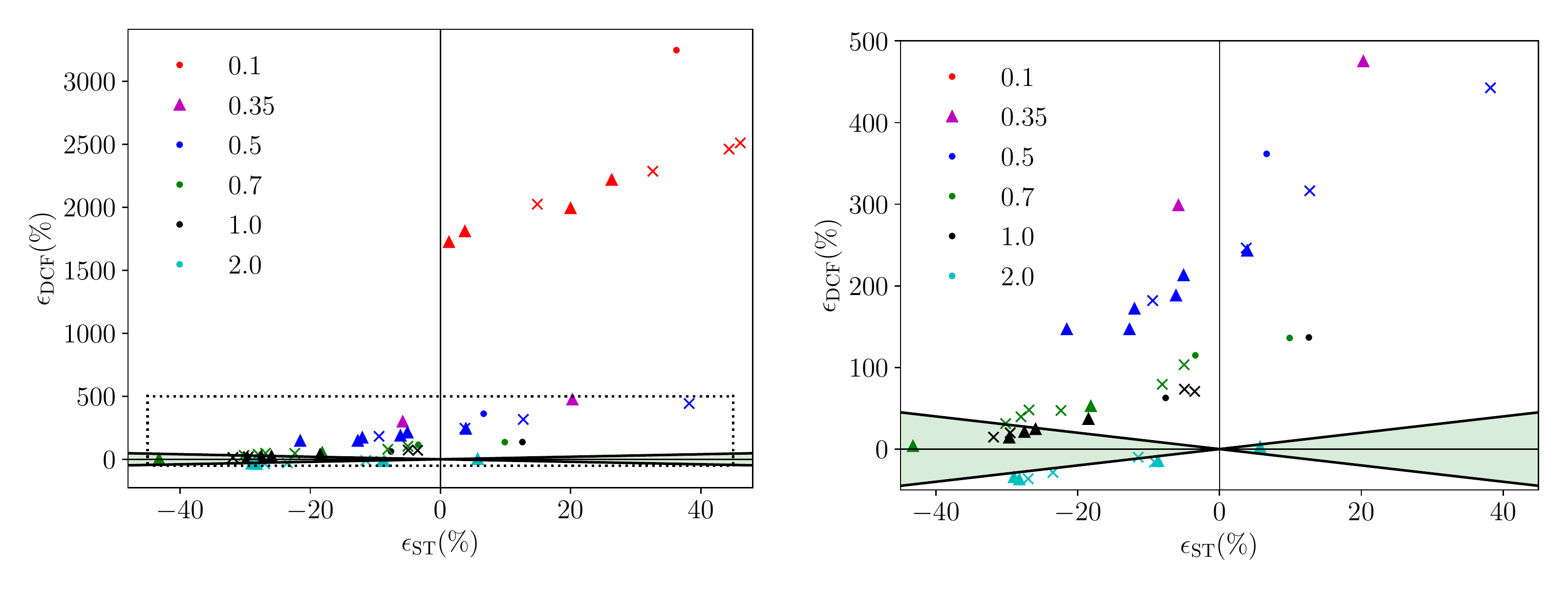}
        \caption{\textbf{Left panel:} Relative error of the DCF method versus the relative error of the ST method. Colors correspond to different \MA\ as shown in the legend. Dots correspond to simulations with $M_{s}<1$, "x" to $1 < M_{s} \leq 4$, and triangles to $M_{s}>4$. The black dotted box marks the zoomed in region shown in the right panel. \textbf{Right panel:} Zoomed region of the left panel.}
     \label{fig:error_dcf_st}
\end{figure*}

\section{Applying the DCF and ST methods in synthetic data}
\label{sec:methods_application}

In this section, we explore the accuracy of the DCF and ST methods in estimating the magnetic field strength. To this end, 
we created synthetic data from every simulation and applied the DCF (with an optimization factor $f=0.5$) and ST methods. Both methods can be significantly affected when the magnetic field is pointing towards the observer since the LOS angle fluctuations induces extra dispersion in \dangles\ \citep{ostriker_2001, falceta_2008, hensley_2019}. For this reason, we assumed that \Bord\ lies completely in the POS.

To obtain an estimate of $\delta u \sim \sqrt{\langle \delta u_\perp^2\rangle} \sim \sigma_{turb}$, we created spectroscopic data in the form of a position-position-velocity (PPV) cube, as in \cite{miville_2003}. We used the following equation,
\begin{equation}
    \label{eq:velocity_cubes}
    I_{v}(x, y, v) = \sum_{LOS} \frac{\rho(x, y, z) }{\sqrt{2\pi} \sigma(x, y, z)} {\rm exp} \left [ -\frac{ \left ( v_{los}(x, y, z) - v \right )^{2} } {2\sigma(x, y, z)^{2}} \right ],
\end{equation}
where $v_{los}(x, y, z)$ is the LOS velocity component and $v$ is the central velocity of each velocity channel. This equation assumes optically thin emission. In Eq.~(\ref{eq:velocity_cubes}), $\sigma(x,y,z)$ is due to thermal broadening and is equal to $\sqrt{k_{B}T/m}$, where $k_{B}$ is the Boltzmann constant.
We then fitted Gaussian profiles to every $I_{v}$ spectrum and derived an "observed" line spread ($\sigma_{obs}$) as a free parameter of the fitting. We computed the turbulent velocity by subtracting in quadrature the thermal broadening,
\begin{equation}
    \sigma_{turb} = \sqrt{\sigma_{obs}^{2} - \frac{k_{B}T}{m} }.
\end{equation}

To obtain an estimate of $\delta \theta$, we applied the formalism of Sect.~\ref{subsec:Bfield_stokes} (Eqs.~\ref{eq:Stokes_Q} and \ref{eq:Stokes_U}) to calculate the Stokes parameters for each LOS, we estimated polarization angles through $\tan2\theta = U/2Q$, and calculated the dispersion of $\theta$ over the entire cloud through $\delta \theta = \sqrt{\langle \theta^2 \rangle_{2D}}$ (taking $\theta=0$ in the direction of the mean magnetic field). 

We then applied the DCF method with $f=0.5$, and the ST method, by inserting the synthetic $\delta u$ and \dangles\ in Eqs.~(\ref{eq:dcf_eq}) and ~(\ref{eq:st_eq}), respectively. We divided both equations with $\sqrt{4 \pi \rho_{0}}$ in order to derive the estimated magnetic field strength in Alfvénic speed units. Finally, we compared the estimated Alfvénic speed ($V_{A}^{est}$) with the actual value ($V_{A}^{true}$) of each simulation and computed their relative error as,
\begin{equation}
    \epsilon (\%) = 100 \frac{V_{A}^{true} - V_{A}^{est}}{V_{A}^{true}}.
\end{equation}

In Fig.~\ref{fig:error_dcf_st} we show the relative error of DCF  ($\rm{\epsilon_{DCF}}$) versus ST ($\rm{\epsilon_{ST}}$). Red color is used for points when $M_{A}=0.1$, magenta for $M_{A}=0.35$, blue for $M_{A}=0.5$, green for $M_{A}=0.7$, black for $M_{A}=1.0$ and cyan for $M_{A}=2.0$. We did not include models with $M_{A} =2.0$ and solenoidal forcing, since the polarization angle distribution of these models is uniform, and thus uninformative. We discuss the effect of forcing of these simulations in more detail in Sect.~\ref{subsec:app_methods_forcing}. The green shaded region corresponds to a smaller error for DCF than for ST. 

We find that DCF is extremely inaccurate at low $M_{A} \leq 0.5$: when the method fails, it fails by factors of several to tens. The accuracy of the method is improved for $M_{A} \geq 1.0$, as expected, since it is in this regime for which the value of $f$ we are using here ($f=0.5$) has been optimized  \citep{ostriker_2001, heitsch_2001, padoan_2001}. The overall trend of this figure is consistent with Fig.~\ref{fig:energetics_ratios}. The DCF method estimates are systematically biased towards larger values, because the kinetic energy of the cloud is much larger than the magnetic fluctuations, even for models with $M_{s} =$ 0.5 and 0.7. The incomressible approximation employed by DCF is reached when $M_{s}$ tends to zero, but even for weakly compressible flows (e.g., $M_{s}\approx0.1$) compressible terms can dominate the dynamics \citep{Bhattacharjee_1998}.
Only in trans/super-Alfvénic cases DCF starts yielding reasonable estimates, because in this regime the contribution of the $\delta B^{2}/(8\pi)$ term in the energy increases. 

In contrast, the error of the ST method remains low and uniform across the different models, because the kinetic energy remains comparable to the \CrossTerm\ fluctuations over a wide range of \MA\ (Fig.~\ref{fig:energetics_ratios}). The ST method overestimates the magnetic field strength at $M_{A} = 0.1$, while at $M_{A}=0.5$ a transition happens. In the latter case half of the measurements overestimate the magnetic field strength, while the rest underestimate it. The underestimation is more prominent at large \MS, since the deviation between the estimated turbulent velocity and the true one becomes larger (Table~\ref{table:sims_results}). ST systematically underestimates the magnetic field strength at models with $M_{A} \geq 0.7$. However, in all cases the error of ST is lower than 50\%. 

\begin{figure*}
    \centering
    \includegraphics[width=\hsize]{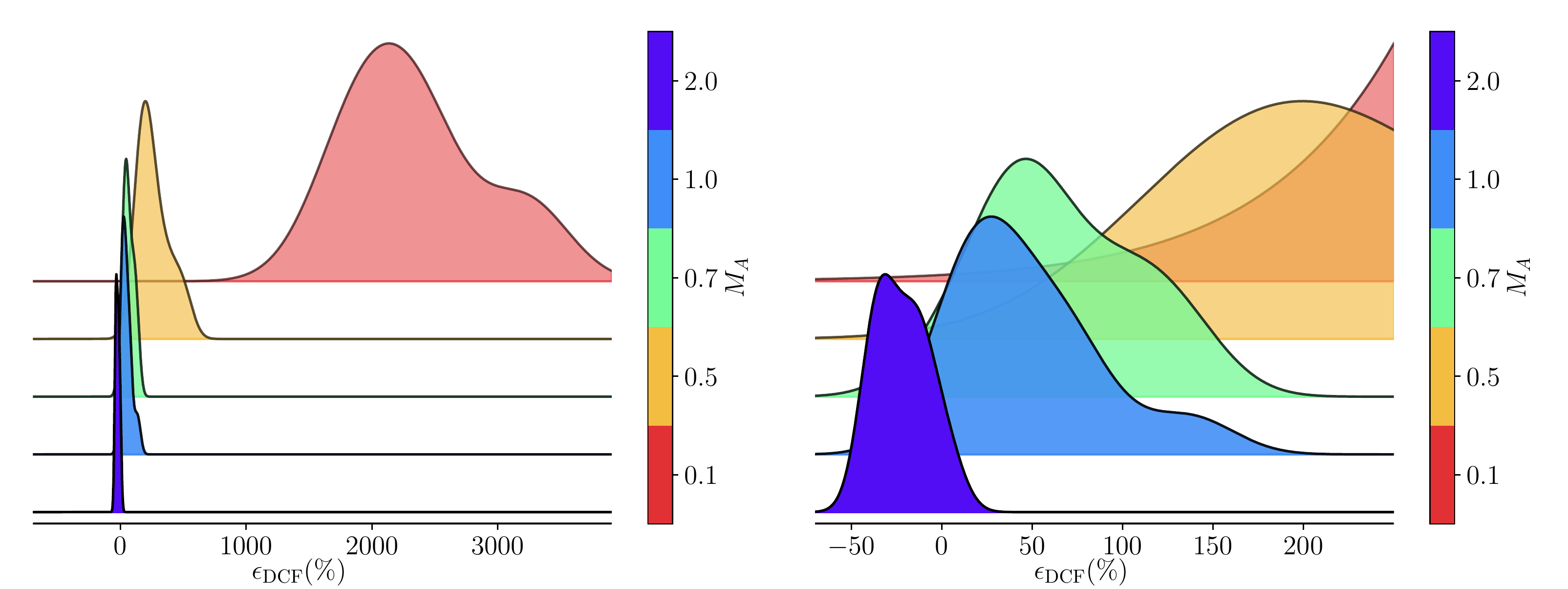}
    \caption{\textbf{Left panel:} Kernel density estimation of $\epsilon_{\rm{DCF}}$ for different \MA\ simulations. \textbf{Right panel:} Zoomed in region of the left panel.}
    \label{fig:kde_dcf}
\end{figure*}

\begin{figure}
    \centering
    \includegraphics[width=\hsize]{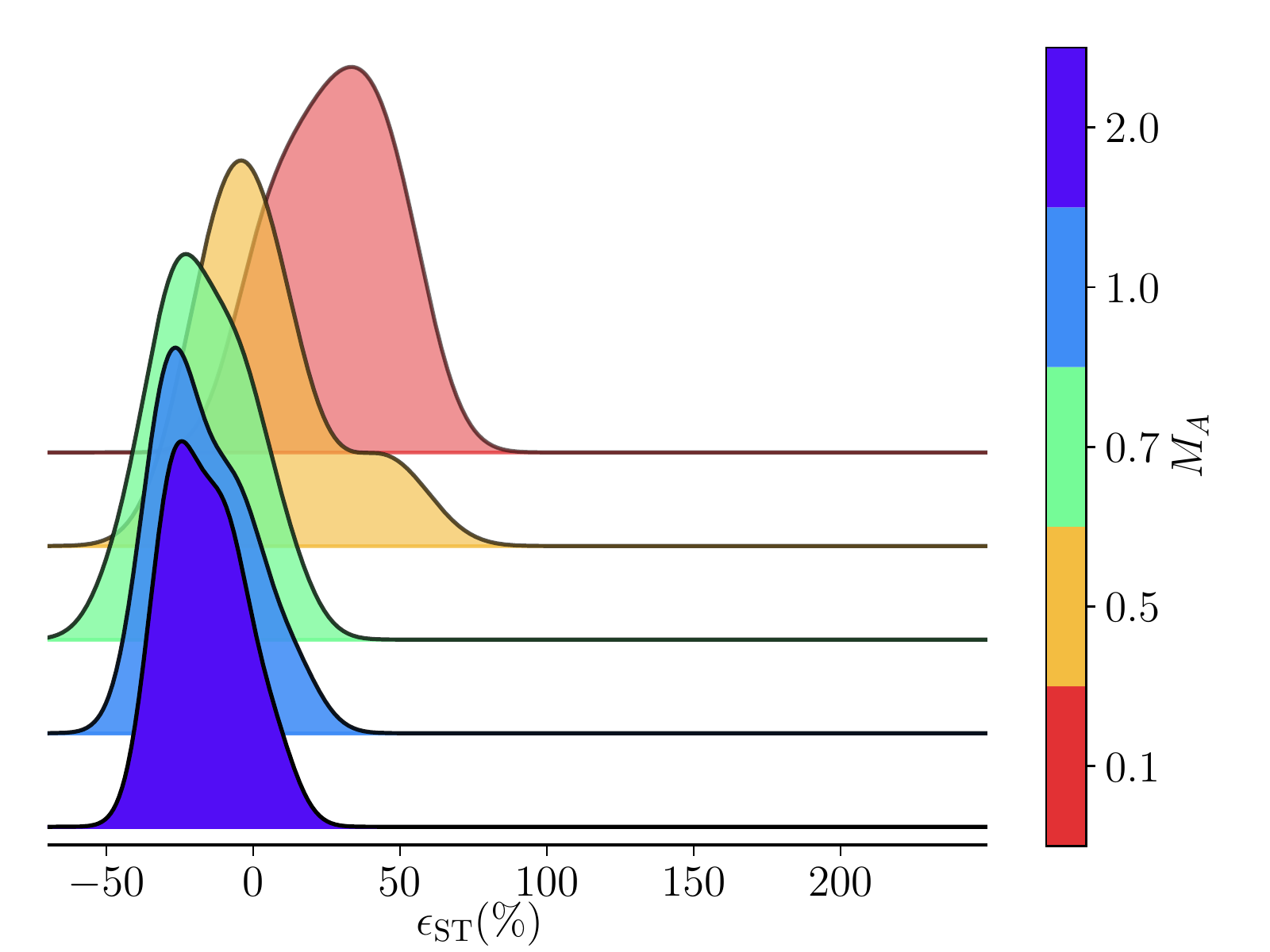}
    \caption{Kernel density estimation of $\epsilon_{\rm{ST}}$ of simulations with different \MA.}
    \label{fig:kde_st}
\end{figure}

\subsection{Statistical properties of \edcf\ and \est.}

In Fig.~\ref{fig:kde_dcf} we show the error distribution for the DCF method, \edcf,  at different \MA, using kernel density estimation. The colorbar indicates the \MA\ of each distribution. Red, yellow, green, cyan and blue correspond to simulations with \MA\ of 0.1, 0.5, 0.7, 1.0 and 2.0, respectively. It is evident that as \MA\ increases \edcf\ decreases. Distributions become more symmetric and narrow at larger \MA\ since the properties of these models are more isotropic. The \edcf\ mean, median and standard deviation of all the models with $M_{A} \geq 0.7$ are $34\%$, $28\%$ and $49\%$ respectively. This shows that the method estimates are systematically biased towards large values and the distribution is skewed to positive values. We also computed the same statistical quantities for the absolute values of \edcf\ for the same models. The mean, median and standard deviation of $|$\edcf$|$ is $47\%$, $37\%$ and $37\%$ respectively. DCF is completely inaccurate at lower \MA, and hence we do not report any statistics for these models.

The \est\ distributions at different \MA\ are shown in Fig.~\ref{fig:kde_st}. Colors are the same as in Fig.~\ref{fig:kde_dcf}. The ST estimates are biased towards larger values at $M_{A}=0.1$, while at $M_{A} \geq 0.7$ the method estimates are systematically biased towards smaller values. The peak of the $M_{A}=0.5$ distribution is close to zero and the probability of overestimating and underestimating the magnetic field strength is equal there. Distributions become more isotropic at larger \MA, which, similarly to \edcf\, happens because the turbulent properties of these models are more isotropic. The \est\ mean, median and standard deviation of all the models are $-2\%$, $-6\%$ and $24\%$ respectively and the distribution is close to a Gaussian. The same quantities for $|$\est$|$ are $20\%$, $18\%$ and $14\%$ respectively. The \est\ distribution is more symmetric and peaks close to zero. Positive values are dominated by models with $M_{A}\leq 0.5$, while negative by models with $M_{A} \geq 0.7$.

The right panel of Fig.~\ref{fig:kde_dcf} and Fig.~\ref{fig:kde_st} show the error distributions of the two methods (DCF and ST) on the same scale. 

\subsection{How does forcing affect the polarization data?}
\label{subsec:app_methods_forcing}

\begin{figure}
    \centering
    \includegraphics[width=\hsize]{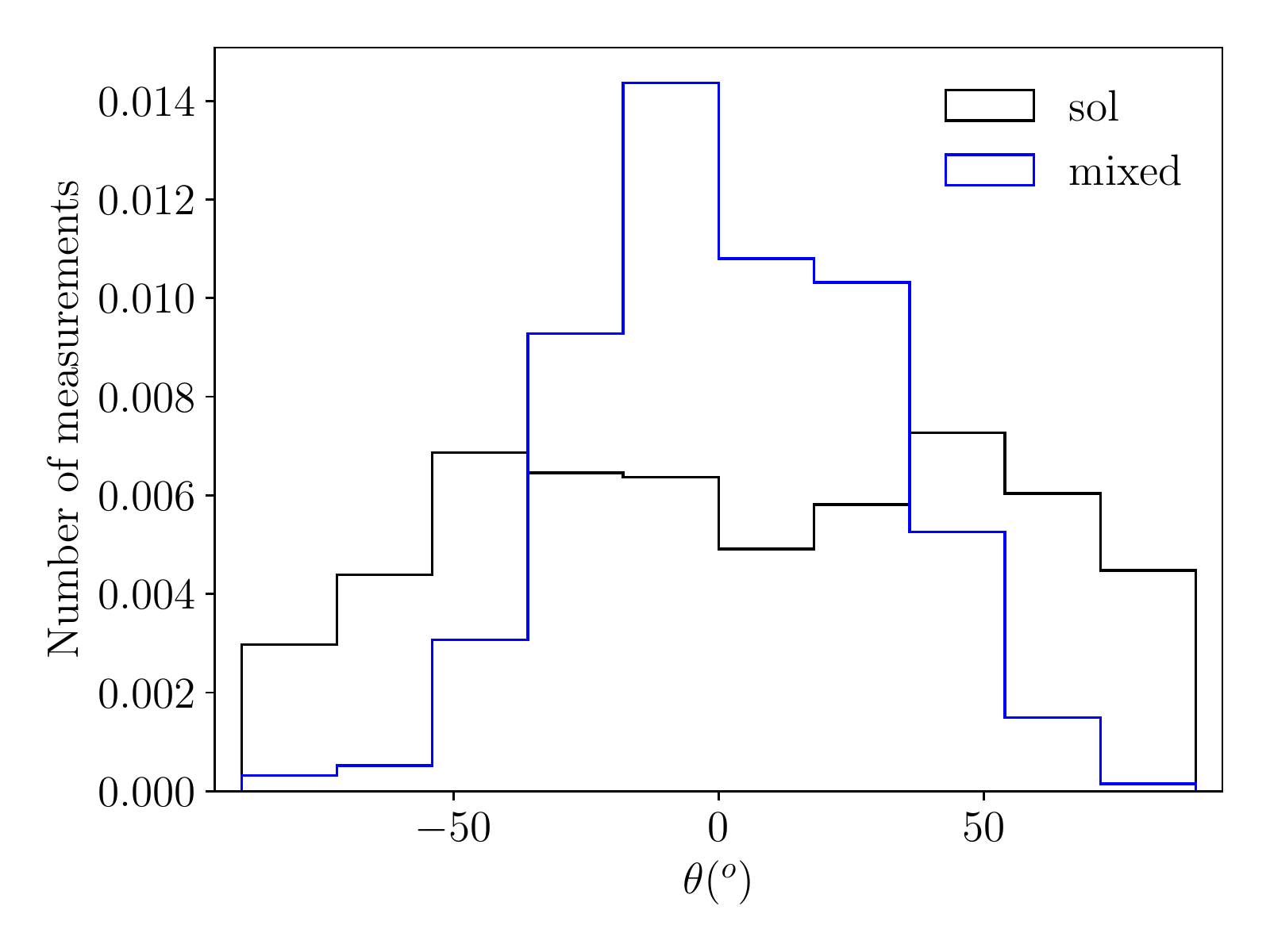}
    \caption{Polarization angle distributions for two different simulations with $M_{A}=2.0$ and $M_{s}=2.0$. Black histogram corresponds to solenoidally driven cloud, while the blue to mixed forcing. Both histograms are normalized by dividing each bin with the total number of measurements. The legend shows the type of forcing.}
    \label{fig:pol_diff}
\end{figure}

The dynamics of a sub-Alfv\'{e}nic cloud are mainly determined by \Bord\ and forcing has a weak role. On the other hand, in super-Alfv\'{e}nic turbulence forcing plays a dominant role in the cloud dynamics (Sect.~\ref{subsec:testing_theory_forcing}). In Sect.~\ref{sec:methods_application} we could not apply the two methods (ST and DCF) in simulations with $M_{A}=2.0$ that were driven solenoidally, but we did apply them to simulations with $M_{A}=2.0$ and mixed forcing. The reason is that forcing has a strong impact on the distribution of polarization angles in super-Alfv\'{e}nic simulations. In Fig.~\ref{fig:pol_diff} we show the normalized polarization angle distributions of simulations with $M_{A}=2.0$ and $M_{s}=2.0$. The black histogram corresponds to a cloud driven solenoidally, while the blue histogram to a cloud driven with a mixture of modes. The black histogram is uniform and characterized by a spread equal to $48\degr$, while the blue histogram has still a well defined mean with a spread equal to $28\degr$. The dispersion in the black histogram (solenoidal driving) is so large it no longer encodes the magnetic fluctuations, due to the limited domain range, \dangles\ $\epsilon~[-90\degr, 90\degr]$.
In this case neither of the two methods can be applied. 
%
%
\section{Conclusions}
\label{sec:conclusions}

The DCF and ST methods have been proposed to estimate the magnetic field strength from dust polarization data. DCF employs the incompressible approximation and infer that $\delta \theta \propto M_{A}$, while ST relaxes this assumption and infers that $\delta \theta \propto M_{A}^{2}$. In this work, we tested both scalings with synthetic data that we produced from ideal-MHD numerical simulations of isothermal clouds without self-gravity, spanning a wide range of \MS\ and \MA\ ($M_{s}~\epsilon~[0.5, 20]$ and $M_{A}~\epsilon~[0.1, 2.0]$). In total we used $26$ different models. We found that the synthetic polarization data can be fit very well with the ST scaling and not with the DCF scaling (Fig.~\ref{fig:dtheta_Ma}). This means that the assumptions and approximations of the DCF method do not hold in compressible turbulence, while the ST assumptions are valid over the entire range of \MA\ studied here.

The major difference between the two methods is that in the DCF energy equation (Eq.~\ref{eq:dcf_energy}) by definition $\vec{\delta B} \cdot \vec{B_{0}} = 0$, since only Alfv\'{e}nic distortions are assumed to be present. On the other hand, ST relax this assumption and consider the more general case of compressible fluctuations where $\vec{\delta B} \cdot \vec{B_{0}} \neq 0$ (Eq.~\ref{eq:st_model}). 

We have explored whether the averaging over the total volume of a cloud can make this term vanish in the energetics, even if locally it is non-zero. We showed that $\langle \vec{\delta B} \cdot \vec{B_{0}} \rangle = 0$ is obtained only if the zero-point of the magnetic "potential energy" is taken to be at $\vec{\delta B}=0$, rather than at the point of minimum potential energy, $\vec{\delta B} = -\vec{\delta B}_{rm max}$. By making a simple analogy with the problem of a bouncing ball in Newtonian gravity, we showed that for compressible fluctuations the correct energy conservation equation is obtained when comparing the kinetic energy with the rms of the \CrossTerm\ in the magnetic energy equation, which then naturally leads to the ST equation for the magnetic field strength (Eq.~\ref{eq:st_model}). We have tested our theoretical arguments with numerical data and found that indeed the rms of \CrossTerm\ compares very well with the kinetic energy in all cases studied, with the exception of super-Alfv\'{e}nic simulations with solenoidal forcing (see Fig.~\ref{fig:energetics_ratios}). On the other hand, when we omit this term, the kinetic energy can be up to two orders of magnitude larger than the magnetic energy. This explains why ST predict the right scaling between \dangles\ - \MA, while DCF do not. This result highly supports the magnetic field fluctuations model of \cite{fedderath_2016} and \cite{beattie_2020}. 

We explored if \CrossTerm\ is imprinted in \dangles. We showed analytically that when $|\vec{B_{0}}| \gg |\vec{\delta B}|$ the zeroth order approximation of \dangles\ refers to perpendicular magnetic field fluctuations and the first order corrections correspond to compressible modes. However, the rms amplitude of parallel and perpendicular fluctuations is in all cases comparable, with deviations smaller than a factor of 2. As a result, \dangles\ provides, indirectly, information about the amplitude of parallel fluctuations even to zeroth order. 

We tested  the accuracy of the two methods in estimating the magnetic field strength. We found that the DCF method with a ``fudge factor'' $f=0.5$ failed completely in clouds with $M_{A} \leq 0.5$ and only started producing reasonable estimates when $M_{A} \gtrsim 0.7$. The lowest errors for DCF were achieved 
for 
trans/super-Alfvénic turbulence, $M_{A} \geq 1.0$, because this is the regime for which the ``fudge factor'' $f=0.5$ we have used was fine-tuned by \cite{ostriker_2001}, \cite{heitsch_2001} and \cite{padoan_2001}. Even in this regime ($M_{A} \geq 0.7$ clouds), the relative error of DCF lied in the range $[-37, 137]\%$. Over the entire $M_A$ range, the error of DCF lied in the range $[-37, 3500]\%$.
The ST method on the other hand gave good results for all $M_A$ examined here, without any fine tuning:  the relative deviation from the true value lied in the range $[-43, 51]\%$ over the entire \MA. We did not find any strong dependence of the accuracy of the methods on \MS. Even in the cases where DCF would outperform ST, the ST method will still provide an adequate estimate of the magnetic field strength, while the reverse is not true. 

\begin{acknowledgements}
We are grateful to Dr. B. Körtgen, Dr. S. Walch, Dr. D. Seifried, and Dr. P. Hennebelle for sharing their simulation data. We would like to thank Dr. E. Ntormousi and Dr. V. Pelgrims for stimulating discussions, N. D. Kylafis and the anonymous referee for valuable comments on the draft. RS would like to thank Dr. K. Christidis for his constant support. This work was supported by the European Research Council (ERC) under the European Unions Horizon 2020 research and innovation programme under grant agreement No. 771282. V. P. acknowledges support from the Foundation of Research and Technology - Hellas Synergy Grants Program through project MagMASim, jointly implemented by the Institute of Astrophysics and the Institute of Applied and Computational Mathematics and by the Hellenic Foundation for Research and Innovation (H.F.R.I.) under the “First Call for H.F.R.I. Research Projects to support Faculty members and Researchers and the procurement of high-cost research equipment grant” (Project 1552 CIRCE). J.~R.~B.~acknowledges financial support from the Australian National University, via the Deakin PhD and Dean's Higher Degree Research (theoretical physics) Scholarships, the Research School of Astronomy and Astrophysics, via the Joan Duffield Research Scholarship and the Australian Government via the Australian Government Research Training Program Fee-Offset Scholarship. 
\end{acknowledgements}

\bibliographystyle{aa}
\bibliography{bibliography}

\begin{table*}
\caption{Simulation properties and methods results}             
\label{table:sims_results}      
\centering          
\begin{tabular}{c c c c c c c c c c c }     
\hline\hline    
\vspace{0.05cm}
Ref & Driving &\MA & \MS & $V_{A, true}$ & $\sigma_{turb}$ & \dangles$(\degr)$ & $V_{A}^{ST}$ & $\rm{\epsilon_{ST}} (\%)$ & $V_{A}^{DCF}$ & $\rm{\epsilon_{DCF}} (\%)$\\ 
(1) & (2) & (3) & (4) & (5) & (6) & (7) & (8) & (9) & (10) & (11) \\ \\
\hline       
   1 & mixed & 0.1 & 0.5  & 5    & 0.36/0.27   & 0.06/0.05   & 7.8/6.8     & 55/36        & 165/167     & 3210/3243 \\ \vspace{0.2cm}
   1 & mixed & 0.1 & 2.0  & 20   & 1.47/1.52   & 0.09/0.09   & 26.4/27.1   & 32.3/35      & 476/480     & 2280/2302 \\ \vspace{0.2cm}
   1 & mixed & 0.1 & 4.0  & 40   & 3.25/2.43   & 0.09/0.08   & 57.8/44.9   & 32.3/12.3    & 1034/830    & 2485/1977         \\ \vspace{0.2cm}
   1 & mixed & 0.1 & 10.0 & 100  & 5.63/6.73   & 0.08/0.09   & 103.8/117.5 & 3.8/17.5     & 1910/2049   & 1810/1949 \\ \vspace{0.2cm}
   1 & mixed & 0.1 & 20.0 & 200  & 13.76/11.25 & 0.09/0.09   & 252.6/202.6 & 26.3/1.3     & 4636/3648   & 2218/1724\\ \vspace{0.2cm}
   1 & mixed & 0.5 & 0.5  & 1    & 0.33/0.27   & 1.73/1.53   & 1.5/1.1     & 51.2/6.5     & 6.1/4.6     & 514/361 \\ \vspace{0.2cm}
   1 & mixed & 0.5 & 2.0  & 4    & 1.15/141    & 2.96/1.86   & 3.6/5.5     & -9.6/38.2    & 11.3/21.7   & 181/442\\ \vspace{0.2cm}
   1 & mixed & 0.5 & 4.0  & 8    & 1.98/2.44   & 2.57/2.10   & 8.3/9.0     & 3.7/12.7     & 27.7/33.3   & 246/316\\ \vspace{0.2cm}
   1 & mixed & 0.5 & 10.0 & 20   & 6.11/5.76   & 3.04/2.63   & 18.8/19.0   & -6.1/-5.1    & 57.6/62.6   & 188/213\\ \vspace{0.2cm}
   1 & mixed & 0.5 & 20.0 & 40   & 8.92/13.01  & 3.00/2.63   & 35/2/41.5   & -12.0/3.9    & 109/137     & 172/243\\ \vspace{0.2cm}
   1 & mixed & 1   & 0.5  & 0.5  & 0.26/0.27   & 9.24/6.48   & 0.46/0.56   & -7.6/12.7    & 0.8/1.19    & 63/137\\ \vspace{0.2cm}
   1 & mixed & 1   & 2.0  & 2    & 0.83/1.09   & 9.89/9.16   & 1.41/1.9    & -29.5/-3.5   & 2.4/3.41    & 19.9/70.6\\ \vspace{0.2cm}
   1 & mixed & 1   & 4.0  & 4    & 1,62/2,08   & 10.12/8.61  & 2.7/3.8     & -31.9/-4.9   & 4.6/6.9     & 14.6/73.4\\ \vspace{0.2cm}
   1 & mixed & 1   & 10.0 & 10   & 4.33/4.41   & 10.86/10.14 & 7.0/7.4     & -29.67/-26.0 & 11.4/12.5   & 14.2/24.4\\ \vspace{0.2cm}
   1 & mixed & 1   & 20.0 & 20   & 8.68/9.70   & 10.27/10.14 & 14.5/16.3   & -27.5/-18.5  & 24.2/27.4   & 21/37.0\\ \vspace{0.2cm}
   1 & mixed & 2.0 & 2.0  & 1    & 0.82/0.87   & 32.91/27.65 & 0.76/0.88   & -23.5/-11.5  & 0.7/0.9     & -28.6/-9.9\\ \vspace{0.2cm}
   1 & mixed & 2.0 & 4.0  & 2    & 1.68/1.96   & 37.77/33.43 & 1.5/1.8     & -27.0/-9.2   & 1.3/1.7     & -36.4/-15.9\\ \vspace{0.2cm}
   1 & mixed & 2.0 & 10.0 & 5    & 4.09/4.89   & 37.33/32.82 & 3.6/4.6     & -28.2/-8.7   & 3.1/4.3     & -37.1/-14.7\\ \vspace{0.2cm}
   1 & mixed & 2.0 & 20.0 & 10   & 7.69/10.88  & 33.63/30.36 & 7.1/10.6    & -29.0/5.8    & 6.5/10.3    & -34.5/2.7 \\ \vspace{0.2cm}
   2 & sol   & 0.7 & 0.7  & 0.91 & 0.46./0.39  & 6.21/5.79   & 0.99/0.87   & 9.5/-3.4     & 2.1/2.0     & 135/115 \\ \vspace{0.2cm}
   2 & sol   & 0.7 & 1.0  & 1.60 & 0.58 /0.71  & 7.00/6.25   & 1.2/1.6     & -26.9/-5.0   & 2.4/3.3     & 47.9/103 \\ \vspace{0.2cm}
   2 & sol   & 0.7 & 2.0  & 2.87 & 1.35/1.07   & 7.53/7.63   & 2.6/2.1     & -8.1/-28.0   & 5.2/4.0     & 79.3/39.5  \\ \vspace{0.2cm}
   2 & sol   & 0.7 & 4.0  & 5.09 & 1.89/2.08   & 8.11/7.96   & 3.6/3.9     & -30.2/-22.4  & 6.7/7.5     & 31.3/47.3\\ \vspace{0.2cm}
   2 & sol   & 0.7 & 7.0  & 9.10 & 3.99/2.83   & 8.24/8.58   & 7.4/5.2     & -18.1/-43.2  & 13.9/9.4    & 52.6/3.7\\ \vspace{0.2cm}
   3 & sol   & 0.5 & 7.5  & 2.86 & 0.76/0.71   &  2.6/2.9    & 2.5/2.2     & -12.7/-21.5  & 8.1/7.1     & 185/147 \\ \vspace{0.2cm}
   4 & sol   & 0.35& 10   & 28.6 &  6.4/7.2    & 1.5/1.4     &  26.9/34.4  & -5.8/20.3    & 348/525     & 299/475\\
\hline                  
\end{tabular}
\tablebib{(1)~\citet{beattie_2020}; (2) \citet{burkhart_2009}; (3) \citet{kortgen_2020}; (4) \citet{mocz_2017}}
\tablefoot{$\sigma_{turb}$ is in km/s and \dangles\ in degrees. When estimating $V_{A}^{\rm{ST}}$ and $V_{A}^{\rm{DCF}}$, \dangles\ is used in radians. In Column (2) "sol" refers to solenoidal forcing.}
\end{table*}

\end{document}